\documentclass[12pt,a4paper]{article}
\pdfoutput=1
\usepackage{color}
\usepackage{amssymb,amsmath,bm,bbold}
\usepackage{epsf}
\usepackage{epsfig}
\usepackage{relsize}
\usepackage[dvipsnames]{xcolor}
\usepackage[linktoc=page,bookmarks=false,colorlinks=false,linkbordercolor=RoyalBlue,citebordercolor=ForestGreen,urlbordercolor=CornflowerBlue]{hyperref}
\usepackage{latexsym,mathrsfs,dsfont}
\usepackage[normalem]{ulem} 
\usepackage[compress]{cite}
\usepackage{graphicx}
\usepackage{url}
\usepackage{booktabs}
\usepackage{float}
\usepackage{multirow}
\usepackage{changepage}
\usepackage[hypcap]{caption, subcaption}

\usepackage[titles]{tocloft}

\usepackage{tabularx,colortbl}

\usepackage{fancyhdr}
\advance \headheight by 3.0truept       

\allowdisplaybreaks[1]

\definecolor{red}{cmyk}{0,1,1,0.4}
\definecolor{darkgreen}{rgb}{0.0,0.6,0.0}
\definecolor{cDarkGrey}{RGB}{91,91,91}
\definecolor{cGrey}{RGB}{245,243,238}
\definecolor{cBlue}{RGB}{0,110,191}
\definecolor{cLightBlue}{RGB}{214,237,252}
\definecolor{cRed}{RGB}{196,0,100}
\definecolor{cLightRed}{RGB}{254,222,237}
\definecolor{cGreen}{RGB}{0,166,80}
\definecolor{cLightGreen}{RGB}{254,222,237}
\definecolor{cOrange}{RGB}{221,74,44}
\definecolor{cLightOrange}{RGB}{255,215,210}
\definecolor{cPurple}{RGB}{93,35,125}
\definecolor{cLightPurple}{RGB}{241,230,252}
\definecolor{cYellow}{RGB}{252,191,10}
\definecolor{cISSRBlue}{RGB}{0,111,174}
\definecolor{cISSRGrey}{RGB}{167,169,172}

\newcommand{\beq}{\begin{equation}}
\newcommand{\eeq}{\end{equation}}
\newcommand{\be}{\begin{equation}}
\newcommand{\ee}{\end{equation}}
\newcommand{\bi}{\begin{itemize}}
\newcommand{\ei}{\end{itemize}}
\newcommand{\ba}{\begin{array}}
\newcommand{\ea}{\end{array}}
\newcommand{\beqa}{\begin{eqnarray}}
\newcommand{\eeqa}{\end{eqnarray}}
\newcommand{\bea}{\begin{eqnarray}}
\newcommand{\eea}{\end{eqnarray}}
\newcommand{\beqn}{\begin{eqnarray}}
\newcommand{\eeqn}{\end{eqnarray}}

\newcounter{TODO}

\newcommand{\mev}{\text{MeV}}
\newcommand{\GeV}{\,\text{GeV}}

\newcommand{\vcb}{|V_{cb}|}
\newcommand{\vtd}{|V_{td}|}
\newcommand{\vub}{|V_{ub}|}
\newcommand{\vts}{|V_{ts}|}
\newcommand{\vus}{|V_{us}|}

\newcommand{\epe}{\varepsilon'/\varepsilon}

\def\kpn{K^+\rightarrow\pi^+\nu\bar\nu}
\def\klpn{K_{L}\rightarrow\pi^0\nu\bar\nu}
\def\ksm{K_S\to\mu^+\mu^-}

\newcommand{\IM}{\rm{Im}}

\newcommand{\mdfd}[1]{{\color{magenta}{#1}}}
\newcommand{\eps}{\epsilon}

\newcommand{\BR}{{\cal B}}

\begin{document}

\begin{flushleft}
  {\em Version of \today}
\end{flushleft}

\vspace{-14mm}
\begin{flushright}
  AJB-22-9
\end{flushright}

\medskip

\begin{center}
{\Large\bf\boldmath
    Standard Model Predictions for  Rare
   K and\\ B Decays  without New Physics Infection
}
\\[1.0cm]
{\bf
    Andrzej~J.~Buras
}\\[0.3cm]

{\small
TUM Institute for Advanced Study,
    Lichtenbergstr. 2a, D-85748 Garching, Germany \\[0.2cm]
Physik Department, TU M\"unchen, James-Franck-Stra{\ss}e, D-85748 Garching, Germany
}
\end{center}

\vskip 0.5cm

\begin{abstract}
  \noindent
  The Standard Model (SM) does not contain by definition any new physics (NP)
  contributions to any observable but contains four CKM parameters which are not predicted   by this model. We point out that if these four parameters are determined in   a global fit which includes processes that are infected by NP and therefore by   sources outside the SM, the resulting so-called SM contributions to rare decay branching
  ratios cannot be considered as {{\em genuine}} SM contributions to the latter.
  On the other hand  {{\em genuine}} SM predictions, that are {\em free} from the CKM dependence, can be obtained for suitable ratios of the $K$ and $B$ rare decay branching ratios  to $\Delta M_s$, $\Delta M_d$ and $|\varepsilon_K|$,  all calculated within the SM.   These three observables contain by now only small hadronic uncertainties and are already  well measured so that rather precise    SM predictions for the ratios in question can be obtained. In this context 
  the {\em rapid test} of NP infection in the $\Delta F=2$ sector is provided
  by a $\vcb-\gamma$ plot that involves   
  $\Delta M_s$, $\Delta M_d$, $|\varepsilon_K|$, and the mixing induced   CP-asymmetry $S_{\psi K_S}$. As with the present hadronic matrix elements this test turns out to be {\em negative},
  assuming negligible   NP infection in the $\Delta F=2$ sector and setting  the values of these four observables to
  the experimental  ones, allows to obtain SM predictions for all
  $K$ and $B$ rare decay branching ratios that are most accurate to date and
  as a byproduct to obtain the full CKM matrix on the basis of $\Delta F=2$
  transitions {\em alone.}
  Using this strategy we obtain SM predictions for 26 branching
  ratios for rare semileptonic and leptonic $K$ and $B$ decays with the $\mu^+\mu^-$ pair
  or the $\nu\bar\nu$ pair in the final state.
 Most interesting
  turn out to be the anomalies in the low $q^2$ bin in
 $B^+\to K^+\mu^+\mu^-$ {($4.4\sigma$)} and $B_s\to \phi\mu^+\mu^-$ ($4.8\sigma$).
\end{abstract}

\thispagestyle{empty}
\newpage
\tableofcontents
\newpage
\setcounter{page}{1}

%
%
%
\section{Introduction}
In this decade and the next decade one expects a very significant progress
  in measuring the branching ratios for several rare $K$ and $B$ decays,
  in particular for the decays $\kpn$, $\klpn$, $K_S\to\mu^+\mu^-$,  $B_s\to\mu^+\mu^-$,   $B_d\to\mu^+\mu^-$, $B\to K(K^*)\nu\bar\nu$, $K_L\to\pi^0\ell^+\ell^-$ \cite{Cerri:2018ypt,Bediaga:2018lhg,NA62:2022nah}. Here Belle II, LHCb, NA62, KOTO and later KLEVER at CERN 
  will play very important roles. All these decays are only mildly affected
  by hadronic uncertainties in contrast to several non-leptonic $B$ decays, $K\to\pi\pi$ decays and in particular
  the ratio $\epe$. As the main hadronic uncertainties for these semi-leptonic and leptonic decays   are collected in the formfactors and weak decay constants, further improvements
  by lattice QCD (LQCD) will reduce these uncertainties to the one percent level.
  Similar, in the case of the charm contribution
  to $\kpn$ and $K_L\to\pi^0\ell^+\ell^-$, long distance effects can be separated
  from  short distance effects and calculated by LQCD. This demonstrates
  clearly the importance of LQCD calculations \cite{FlavourLatticeAveragingGroupFLAG:2021npn} in this and coming decades
  \cite{Blum:2022wsz}. For $B$ physics this is also the case of HQET Sum Rules
   \cite{Kirk:2017juj}. But also Chiral Perturbation Theory
  is useful in this context allowing to extract some non-perturbative
  quantities from data on the leading Kaon decays \cite{Cirigliano:2011ny}.

  Of particular interest are also semi-leptonic  decays $B^+\to K^+\mu^+\mu^-$, $B^0\to K^{*0}\mu^+\mu^-$, $B_s\to \phi\mu^+\mu^-$ and $\Lambda_b\to \Lambda\mu^+\mu^-$ which play an important role in the
  analyses of the so-called $B$-physics anomalies. They are not as theoretically
  clean as semi-leptonic decays with neutrinos and in particular leptonic
  decays but they have the advantage of having larger branching ratios so
  that several of them have been already measured with respectable precision.

  As far as short distance QCD and QED calculations within the Standard Model (SM) of the decay branching ratios in question are concerned,  a very significant progress
  in the last thirty years has been achieved.  It is reviewed in  \cite{Buchalla:1995vs,Buras:2011we,Buras:2020xsm,Aebischer:2022vky}. In this manner rather precise formulae for SM
  branching ratios as functions of four CKM parameters \cite{Cabibbo:1963yz,Kobayashi:1973fv} can be written down. It will be useful to choose  these
  parameters as follows\footnote{This choice is {more useful than} the one in which
    $\beta$ is replaced by $\vub$, allowing for much simpler CKM factors
    than in the latter case used e.g. recently in \cite{Altmannshofer:2021uub}.}
\be\label{4CKM}
\boxed{\vus,\qquad \vcb, \qquad \beta, \qquad \gamma}
\ee
with $\beta$ and $\gamma$ being two angles in the Unitarity Triangle (UT). Similarly
SM expressions for  the $\Delta F=2$ observables
\be\label{loop}
\boxed{|\varepsilon_K|,\qquad \Delta M_s,\qquad \Delta M_d, \qquad S_{\psi K_S}}\,
\ee
in terms of the CKM parameters can be written down. Due to the impressive progress
by LQCD and HQET done in the last decade, the  hadronic matrix elements relevant for the latter observables are already known with a high precision. This  is even more the case of short distance QCD contributions for which not only NLO
QCD corrections are known \cite{Buras:1990fn,Herrlich:1993yv,Herrlich:1995hh,Herrlich:1996vf}  but also the NNLO ones \cite{Brod:2011ty,Brod:2010mj,Brod:2019rzc}and the NLO electroweak corrections \cite{Brod:2021qvc,Brod:2022har}. As the experimental precision on 
$|\varepsilon_K|$, $\Delta M_s$ and $\Delta M_d$ is already impressive and
the one on the mixing induced CP-asymmetry $S_{\psi K_S}$, that gives us $\beta$,
will be improved by the LHCb and Belle II collaborations soon, this complex of $\Delta F=2$ observables
is in a much better shape than $\Delta F=1$ transitions if both the status of the experiment and the status of the theory are simultaneously considered.

We have then a multitude of SM expressions for branching ratios, asymmetries
and other observables as functions of only four CKM parameters in (\ref{4CKM})
that are not predicted in the SM. The remaining parameters like $W^\pm$, $Z^0$, quark and lepton masses and gauge coupling constants or Fermi-constant $G_F$
are already known from other measurements. The question then arises whether not
only 
this system of SM equations describes the existing measurements well, but also
what are the SM predictions for rare decay branching ratios measured already
for several $b\to s \mu^+\mu^-$ transitions and to be measured for very rare decays with neutrino pair or charged lepton pair in the final state in this and the next decade.

In the 21st century the common practice is to insert all these equations into a computer code
like the one used by the CKMfitter \cite{Charles:2004jd} and the UTfitter \cite{Bona:2007vi} and more recently popular  Flavio \cite{Straub:2018kue} and HEPfit  \cite{DeBlas:2019ehy} codes among others. In this manner apparently not only the best values for the CKM parameters can be obtained and consistency checks of the SM predictions can be made. Having the CKM parameters at hand, apparently, one can even find  the best SM predictions for various rare decay branching ratios.

While, I fully agree that in this manner a global consistency checks of the SM
can be made, in my view the resulting SM predictions cannot be considered as
 {{\em genuine}} SM
predictions, simply because the values of the CKM parameters and consequently the Unitarity Triangle, obtained in such a
global fit, are likely to depend on possible NP infecting them\footnote{This point has been already made in a short note by the present author \cite{Buras:2022irq} and very
  recently in \cite{DeBruyn:2022zhw} but the solution to this problem suggested in the latter   paper is drastically different from the one proposed here
  that is based on \cite{Buras:2021nns,Buras:2022wpw}. We will comment on it
below.}.
 This is in particular the case if
some inconsistencies in the SM description of the data for certain observables are found and one has to  invoke some models to explain the data. This is
in fact the case of several $b\to s\mu^+\mu^-$ transitions for which data are already available.

Moreover there is another
problem with such global fits at present. It is the persistent tension
  between inclusive and exclusive determinations of $\vcb$ \cite{Bordone:2021oof,FlavourLatticeAveragingGroupFLAG:2021npn}\footnote{The exclusive value
  for $\vcb$ should be considered as preliminary.}
\be\label{TREE}
\vcb_{\rm incl}=42.16(50)\times 10^{-3},\qquad 
\vcb_{\rm excl}=39.21(62)\times 10^{-3},
\ee
which is clearly disturbing because as stressed in \cite{Buras:2021nns}
the SM predictions for rare decay branching ratios and also
$\Delta F=2$ observables in (\ref{loop})
are sensitive functions of $\vcb$. Therefore the question arises  which
of these two values should be used in a global fit if any\footnote{This question applies also to global fits related to the tests of lepton flavour universality violation in which the CKM input only from tree-level decays is used. See
  \cite{Alguero:2022wkd} and references therein.}. As shown
recently in \cite{Buras:2022wpw}, the SM predictions for the branching ratios in question and
$\Delta F=2$ observables are drastically different for these two values of $\vcb$. This problem existed already in 2015 in the context of the widely cited
paper in \cite{Buras:2015qea} as stressed recently in a short note in \cite{Buras:2022irq}.

But this is not the whole story. Many observables involved in the global fits
contain larger hadronic uncertainties than the rare decays listed above and
also larger than the $\Delta F=2$ observables in (\ref{loop})
so that SM predictions for theoretically clean decays are polluted in a global fit by these uncertainties. While such observables can be given a low
weight in the fit, this uncertainty will not be totally removed.

In my view these are  important issues related to global fits that to my knowledge have not been
addressed sufficiently in print by anybody. They  will surely be important 
when in the next years the data on a multitude of branching ratios will improve and the hadronic parameters that are not infected by NP will be better known.
Therefore, the basic question which I want to address here is whether
it is possible to  find accurate SM predictions for rare $K$ and $B$ decays without any NP infection in view of the following three problems which one has to face:
\begin{itemize}
\item
  Several anomalies in semi-leptonic decays, like suppressed $b\to s\mu^+\mu^-$ branching ratios. 
\item
  Significant tensions between inclusive and exclusive determinations of $\vcb$
  implying very large uncertainties in the SM predictions for rare decay branching ratios and making the  use of the values of $\vcb$ from tree-level decays in this context questionable. Moreover, it is not yet excluded that these tensions are caused by NP \cite{Colangelo:2016ymy}.
\item
  Hadronic uncertainties in various well measured observables included in a global fit that
  are often much larger than the ones in rare $K$ and $B$ decays.
\end{itemize}

The present paper suggests a possible solution to these problems and
studies its implications. It is based on the ideas developed in collaboration
with Elena Venturini  \cite{Buras:2021nns,Buras:2022wpw} and
extends them in a significant manner. The short note in \cite{Buras:2022irq}
by the present author, in which some critical comments about the literature
have been made,  can be considered as an overture to the present paper.
{In fact our strategy is consistent with the present pattern of experimental data. While significant NP effects have been found in $\Delta F=1$ processes, none in $\Delta F=2$
  processes. This peculiar situation has been already addressed in 
  the context of $B$ physics anomalies by other authors and we will
add a few additional remarks at the end of our paper. However, in none
of the related papers in the literature the suggestion has been made to use this fact for the determintation of the CKM parameters without NP infection from $\Delta F=2$ observables
alone, so that the strategies developed
in  \cite{Buras:2021nns,Buras:2022wpw} and used extensively in the
present paper open a new route to phenomenology of flavour violating processes,
not only in the SM but also beyond it.}

The outline of our paper is as follows.
In Section~\ref{sec:2} we will briefly explain why the SM predictions for rare decays resulting from a  global fit cannot be considered as  {{genuine}} SM predictions unless a careful choice of the observables included in the fit is made.
In Section~\ref{sec:3} I will argue that the strategy developed recently
in collaboration with Elena Venturini  \cite{Buras:2021nns,Buras:2022wpw}
is  presently the most efficient method for obtaining CKM-independent SM predictions for various suitable ratios of rare decay branching ratios
to the $\Delta F=2$ observables in (\ref{loop}).
In Section~\ref{sec:3a}
we address the issue of predicting SM  branching ratios themselves. To this end we make the assumption that NP
contributions to $\Delta F=2$ observables are negligible which is motivated by a {\em negative} rapid test that shows
a very consistent description of the very precise
experimental data on these observables within the SM. This is in addition
supported by a new CKM free SM relation (\ref{RNEW}) between the four $\Delta F=2$ observables
in (\ref{loop}) that is in a very good agreement with the data.

Setting  the values of $\Delta F=2$ observables to
  their experimental  values and using the CKM-independent ratios found above, allows to obtain SM predictions for all very rare 
  $K$ and $B$ branching ratios that are most accurate to date \cite{Buras:2021nns,Buras:2022wpw}.
  Another bonus of this strategy is the determination of the CKM parameters
  from $\Delta F=2$ processes {\em alone}, that allows in turn to make accurate predictions
  for  a number   of $\vcb$-independent ratios that depend on $\beta$ and $\gamma$ \cite{Buras:2021nns}. {In Section~\ref{sec:3b} using these CKM parameters we find  SM predictions for the branching ratios
  of $B^+\to K^+\mu^+\mu^-$, $B^0\to K^{*0}\mu^+\mu^-$, $B_s\to \phi\mu^+\mu^-$ and $\Lambda_b\to \Lambda\mu^+\mu^-$ and in Section~\ref{sec:3c}}
  SM predictions for several $B_s$ decays with $\nu\bar\nu$ in the final state
  are presented.

  {However, it should be stressed that the predictions in Sections~\ref{sec:3b}
  and ~\ref{sec:3c} go beyond the main strategy of removing CKM parameters
    from the analyses and in Section~\ref{sec:3d} we repeat the calculation
    of the decays considered in Sections~\ref{sec:3b}
    and ~\ref{sec:3c} by eliminating $\vts$ with the help of $\Delta M_s$
      and setting its value to the experimental one. As expected we
      find very similar results but they are more stable under future
      modifications of $\vts$ due to possible changes in
      non-perturbative parameters in the $\Delta F=2$ system beyond those
      relevant for $\Delta M_s$.}

  In our view the strategies presented here allow to assess better
  the pulls in individual branching ratios than it is possible in a global
  fit, simply because the assumption of the absence of NP is made only
  in $\Delta F=2$ observables which constitute a
  subset of observables used in global fits. As within this subset
  no NP is presently required to describe the data, the resulting 
  SM predictions for rare decays are likely to be free from NP infection.
  In Section~\ref{EXHYB} we make a few comments on the so-called EXCLUSIVE
  and HYBRID scenarios based on tree-level decays and considered already in detail in \cite{Buras:2022wpw}. They could be realized one day if the experts
  agree on the unique values of $\vcb$ and $\vub$.
  In Section~\ref{sec:5a} we outline the strategy for finding footprints of NP before
  one    starts using computer codes.
  A brief summary and an  outlook are given in Section~\ref{sec:4}.

  Before we start I would like to stress that I am making here a point
  which I hope will be taken seriously by all flavour practitioners, not only
  by global fitters. If one does not want to
  face the tensions in the determination of $\vcb$ and $\vub$ through tree-level
  decays, the $\Delta F=2$ route is presently the only one possible. The
  tree-level route explored recently in  \cite{DeBruyn:2022zhw} in detail is presently   much harder and is in my view not as transparent as the  $\Delta F=2$ route \cite{Buras:2003td,Buras:2021nns,Buras:2022wpw} followed here.
  In particular it did not lead yet to unique values of the CKM parameters
  because of the tensions between the exclusive and inclusive determinations
  of $\vub$ and $\vcb$.

  In fact the basic idea, beyond the removal of the CKM dependence with the help of suitable ratios \cite{Buras:2003td,Buras:2021nns,Buras:2022wpw} and subsequently using {\em only} $\Delta F=2$ observables to find CKM parameters,
  can be formulated in a simple manner as follows. Imagine the $\Delta F=2$  archipelago
consisting of the four $\Delta F=2$ observables in (\ref{loop}). They
can be precisely measured and the relevant hadronic matrix elements
can be precisely calculated by using LQCD and HQET Sum Rules.
This is sufficient to determine CKM parameters using the SM expressions for these observables finding that this model can consistently describe them  \cite{Buras:2022wpw}. But LQCD and HQET experts can calculate all non-perturbative quantities like weak decay constants, formfactors, hadronic matrix elements etc. so that SM predictions for quantities outside the $\Delta F=2$  archipelago can be made. Comparing these
predictions with experiments outside this archipelago one can find out whether
there are phenomena that cannot be described by the SM.

To my knowledge
there is no analysis in the literature, except for \cite{Buras:2021nns,Buras:2022wpw}, that made SM predictions for rare decay
observables using this simple strategy. In the present paper we extend this
stategy to several decays not considered in \cite{Buras:2021nns,Buras:2022wpw},
in particular those in which anomalies have been found.

The numerous results following from this strategy are 
  presented in Tables~\ref{tab:SMBRBV2}-\ref{tab:nunubar}, in the 
  formulae (\ref{eq:R0})-(\ref{R12ab}) and (\ref{R13new})-(\ref{R14newf}).
Some of them can be already
compared with existing data and many will be compared with improved experimental data which will be available in this decade.

\section{New Physics Infected Standard Model Predictions}\label{sec:2}
Let us consider a global SM fit which exposes some  deficiencies of this model
summarized as anomalies. There are several anomalies in various decays observed in the data, 
in particular {in semi-leptonic $B$ decays with  a number of branching ratios found below
SM predictions, }
 the $(g-2)_\mu$ anomaly and the Cabibbo anomaly among others as reviewed recently in
 \cite{Crivellin:2022qcj}. There is some NP hidden behind these anomalies.
 The most prominent candidates for this NP are presently  the leptoquarks, vector-like quarks and $Z^\prime$. Even if in a SM global fit all these NP contributions are
 set to zero, in order to see the problematic it is useful to include
 them in a specific BSM model with the goal  to remove these anomalies. The  branching ratio for
a specific rare decay resulting from such a fit has the general structure
\be
\mathcal{B}(\text{Decay})= \mathcal{B}(\text{Decay})^i_{\text{SM}}+
\mathcal{B}(\text{Decay})^i_{\text{BSM}}
\ee
in the case of no intereference between SM and BSM contributions or for
decay amplitudes 
\be
\mathcal{A}(\text{Decay})= \mathcal{A}(\text{Decay})^i_{\text{SM}}+
\mathcal{A}(\text{Decay})^i_{\text{BSM}}
\ee
in the case of the intereferences between SM and NP contributions. The index $i$ distinguishes different BSM
scenarios. The dependence of the SM part on BSM scenario considered enters
exclusively through CKM parameters that in a global fit are affected
by NP in a given BSM scenario. Dependently on the BSM scenario,
different SM prediction result for a given  decay which is at least for me  a problem. In the SM there is no NP by definition and there must be  a unique SM prediction for a given decay that can be directly compared with experiment.

It could be that for some flavour physicists, who only worked in BSM scenarios
and never calculated NLO and NNLO QCD corrections to any decay, this is not a problem.
However, for the present author and many of his collaborators as well as other
flavour theorists, who spent years
calculating higher order QCD corrections to many rare decays, with the goal to find
precise  {{genuine}} SM predictions for various observables, it is a problem and should be a problem. 
But to me the important question is also whether in a global fit the values in (\ref{HYBRID}) should be taken into account or not.
Such questions are avoided in the strategy of \cite{Buras:2003td} and
\cite{Buras:2021nns,Buras:2022wpw} because $\vcb$ is eliminated from the
start.

This should also be a problem for LQCD experts  who for hadronic matrix elements
relevant for $\Delta M_s$, $\Delta M_d$ and $\varepsilon_K$, weak decay constants and formfactors achieved for some of them the accuracy in the ballpark of $1\%$.

In order to exhibit this problematic in explicit terms it is useful to
quote the determination of the CKM elements $\vcb$ and $\vub$ from most important flavour changing
loop transitions that have been measured, that is meson oscillations and rare
$b$ hadron decays, including those  that show
anomalous behaviour \cite{Altmannshofer:2021uub}
\be\label{alt}
\vcb_{\text{loop}}=(41.75\pm 0.76)\times 10^{-3}, \qquad \vub_{\text{loop}}=(3.71\pm 0.16)\times 10^{-3}.
\ee
The authors of \cite{Altmannshofer:2021uub} stressed that these values
should not be used to obtain SM predictions and we fully agree
with them. But in order to assess the size of $B$-physics  anomalies properly,
we would like to make SM predictions that are not infected by NP.
We will soon see that $\vcb$ in (\ref{alt}) is indeed infected by NP.

It is probably a good place to comment on the very recent paper in \cite{DeBruyn:2022zhw} in which the authors emphasized that in the process of the determination
of the CKM parameters care should be taken to avoid observables that are likely to be  affected by NP contributions, in particular the $\Delta F=2$ observables
which are key observables for the determination of the CKM parameters in the
present paper and also in \cite{Buras:2021nns,Buras:2022wpw}.
Trying to avoid  the $\Delta F=2$ observables in their determination
of the CKM parameters as much as possible they were forced to consider various scenarios for the $\vcb$ and $\vub$ parameters that suffer from the tensions
mentioned above. The fact that exclusive and inclusive values of
these parameters imply very different results for rare $K$ and $B$ decays
as well as for the $\Delta F=2$ observables in (\ref{loop}) has been already
presented earlier in \cite{Buras:2022wpw}, but the authors of \cite{DeBruyn:2022zhw}
gave  additional insights in this problematic. Moreover,
they study the issue of the $\gamma$ determinations in non-leptonic $B$
decays which will also be important for the tests of our strategy.

Our strategy is much simpler and drastically different from  the one of 
\cite{DeBruyn:2022zhw} and the common prejudice, also expressed by the latter
authors, that $\Delta F=2$ observables are likely to
be affected by NP. Presently nobody can claim that these observables are affected
by NP. Assuming then, in contrast to \cite{DeBruyn:2022zhw}, that NP contributions
to $\Delta F=2$ observables are negligible allows not only to avoid
tensions in $\vcb$ and $\vub$ determinations that have important implications
on SM predictions for flavour observables \cite{Buras:2021nns,Buras:2022wpw}.
It  also allows to determine uniquely and
precisely CKM parameters so that various scenarios for them presented
in  \cite{DeBruyn:2022zhw} as a result of the tensions in question 
can be avoided.

Needless to say I find the analysis in  \cite{DeBruyn:2022zhw} interesting and very informative. It will certainly be useful if clear signals of NP will
be identified in $\Delta F=2$ observables.
Next years will tell us whether their strategy or our strategy is more
successful in obtaining SM predictions for a multitude of flavour observables.

\section{SM Predictions for CKM-independent Ratios}\label{sec:3}

The only method known to me that allows presently to find SM predictions
for rare $K$ and $B$ decays
without any NP infection is to 
consider suitable ratios of rare decay branching ratios calculated
in the SM to the first three $\Delta F=2$ observables in (\ref{loop}), calculated also
in the SM, so that the CKM dependence is eliminated as much as possible,
in particular the one on $\vcb$ completely. This proposal in the case
of $B_{s,d}\to \mu^+\mu^-$ decays, that in fact works for all $B$-decays
governed by $\vtd$ and $\vts$ couplings, goes back to 2003 in which
the following CKM-independent SM ratios have been proposed \cite{Buras:2003td}
\be\label{CMFV61}
\boxed{R_q=\frac{\overline{\mathcal{B}}(B_q\to\mu^+\mu^-)}{\Delta M_q}= 4.291\times 10^{-10}\ \frac{\tau_{B_q}}{\hat B_q}\frac{(Y_0(x_t))^2}{S_0(x_t)},\qquad q=d,s\,,}
\ee
with $Y_0$ and $S_0$ known one loop $m_t$-dependent functions. The parameters
$\hat B_q$ are known already with good precision from LQCD  \cite{Dowdall:2019bea}. The ``bar'' on the branching ratios takes into account the $\Delta\Gamma_q$
effects that are only relevant for $B_s\to\mu^+\mu^-$ \cite{deBruyn:2012wk}.

Recently this method has been generalized to rare Kaon decays. Presently the most interesting $\vcb$-independent ratios in this case read \cite{Buras:2021nns,Buras:2022wpw} \footnote{The nominal value of $\gamma$ in these expression as used
  in \cite{Buras:2021nns,Buras:2022wpw} differs from $\gamma=64.6^\circ$ used by us in subsequent papers  but inserting the latter 
  has practically no impact on the numerical coefficients in these ratios.}
\be\label{R11}
  \boxed{R_{11}(\beta,\gamma)=\frac{\mathcal{B}(\kpn)}{|\varepsilon_K|^{0.82}}=(1.31\pm0.05)\times 10^{-8}{\left(\frac{\sin\gamma}{\sin 67^\circ}\right)^{0.015}\left(\frac{\sin 22.2^\circ}{\sin \beta}\right)^{0.71},  }            }
  \ee
  \be\label{R12a}
\boxed{R_{12}(\beta,\gamma)=\frac{\mathcal{B}(\klpn)}{|\varepsilon_K|^{1.18}}=(3.87\pm0.06)\times 10^{-8}
    {\left(\frac{\sin\gamma}{\sin 67^\circ}\right)^{0.03}\left(\frac{\sin\beta}{\sin 22.2^\circ}\right)^{0.9{8}},}}
  \ee
  where the ratios $R_q$, $R_{11}$ and $R_{12}$ belong to the set of 16 $\vcb$-independent ratios proposed in \cite{Buras:2021nns}. We will encounter them
  in Section~\ref{ratios}.

  It should be stressed that these ratios are valid {\em only} within the SM.
It should also be noted  that
the only relevant CKM parameter in these $\vcb$-independent ratios
 is the UT angle $\beta$ and this is the reason why we need the mixing induced CP-asymmetry $S_{\psi K_S}$ to obtain predictions for $\kpn$ and $\klpn$.
 While $\gamma$ also enters these expressions, its impact on final results is practically irrelevant. This is still another advantage of this strategy
 over global fits in addition to the independence of $\vcb$ because
 while $\beta$ is already rather precisely known, this is not the case for $\gamma$:
  \be\label{betagamma}
  \boxed{\beta=(22.2\pm 0.7)^\circ, \qquad  \gamma = (63.8^{+3.5}_{-3.7})^\circ \,.} \ee
  Here the value for $\gamma$ is the most recent one from the LHCb which updates
  the one in \cite{LHCb:2021dcr} $(65.4^{+3.8}_{-4.2})^\circ$.
  However, as we will see below our strategy will allow the determination of
  $\gamma$ that is significantly more precise than this one and in full agreement with the LHCb value above.
  
  Yet, even if  in the coming years the determination of $\gamma$ by the LHCb and Belle II collaboratios will be significantly improved and this will certainly have an
  impact on global fits, this will have practically no impact on the 
  SM predictions for the four ratios listed above. On the other hand  the improvement on the measurement of $\beta$ will play more important role for
  $R_{11}$ and $R_{12}$ and thereby also for $\kpn$ and $\klpn$ 
decreasing the uncertainty in the SM predictions
  for both decays. For $\kpn$ further improvement will be obtained
  by reducing the uncertainty in long distance charm contribution through
  LQCD computations \cite{Blum:2022wsz}. Then the uncertainty in the numerical
  factor in $R_{11}$ will be further decreased allowing to test the SM in 
   an impressive manner  when the $\kpn$ branching ratio will be measured
  at CERN in this decade with an accuracy of $5\%$.

  Before continuing let us stress again that the results for the ratios $R_q$, $R_{11}$ and $R_{12}$ are only valid in the SM and being practically independent of the CKM
  parameters can be regarded as {genuine} SM predictions for the ratios
  in question.  Except for $\beta$ obtained using SM expression
 \be\label{beta}
 S_{\psi K_S}=\sin(2\beta)=0.699(17)\,
 \ee
 I do not have to know other CKM parameters to obtain the SM predictions listed
 above.

 The experimental values of the $\Delta F=2$ observables in
 (\ref{loop}) are already known with high precision. Once the four
 branching ratios will be experimentally known these four ratios
 will allow a very good test of the SM without any knowledge of the CKM
 parameters except for $\beta$ in the case of $\kpn$ and $\klpn$. We will
 return to other ratios in Section~\ref{ratios}.

 \section{SM Predictions for Rare Decay Branching Ratios}\label{sec:3a}
\subsection{Main Strategy and First Results}

 But this is the story of the ratios. We would like to make one step further and
   obtain SM predictions for branching ratios themselves. The proposal
   of \cite{Buras:2021nns,Buras:2022wpw} is to use in the ratios in
   question the experimental values for the $\Delta F=2$ observables in
   (\ref{loop}) to predict the branching ratios for rare $K$ and $B$ decays.
   There are four arguments for this procedure:
   \begin{itemize}
   \item
     The experimental status of $\Delta F=2$ observables is much better
     than the one of rare decays and their theoretical status is very good.
   \item
     To obtain  SM predictions for branching ratios that are not infected
     by NP the only logical possibility is to assume that SM describes
     properly $\Delta F=2$ observables not allowing them to be infected by
     NP.
   \item
     The latter assumption is supported  by the data on $\Delta F=2$
     observables as pointed out in \cite{Buras:2022wpw} and repeated below. There is presently  no need  for NP contributions to $\Delta F=2$ observables to
     fit the data.
   \item
     There is no other sector of flavour observables that can determine all CKM parameters beyond $\vus$, in particular $\vub$ and $\vcb$, in which the tensions between inclusive and exclusive determinations of the latter can be avoided.
     \end{itemize}

\begin{table}
\centering
\renewcommand{\arraystretch}{1.4}
\resizebox{\columnwidth}{!}{
\begin{tabular}{|ll|l|}
\hline
Decay 
& SM Branching Ratio
& Data
\\
\hline \hline
 $B_s\to\mu^+\mu^-$ &  $(3.78^{+ 0.15}_{-0.10})\cdot 10^{-9}$      &  $(3.45\pm0.29)\cdot 10^{-9}$ \cite{LHCb:2021awg,CMS:2020rox,ATLAS:2020acx,HFLAV:2022pwe} 
\\
 $B_d\to\mu^+\mu^-$ &  ${(1.02^{+ 0.05}_{-0.03})}\ \cdot 10^{-10}$      & $\le 2.05\cdot 10^{-10}$ \cite{LHCb:2021awg}
\\
 $B_s\to\tau^+\tau^-$ &  $(7.94^{+ 0.32}_{-0.21})\cdot 10^{-7}$      &  $\le 6.8\cdot 10^{-3}$ \cite{Aaij:2017xqt}
\\
 $B_d\to\tau^+\tau^-$ &  ${(2.14^{+ 0.10}_{-0.06})}\ \cdot 10^{-8}$      & $\le 2.1\cdot 10^{-3}$ \cite{Aaij:2017xqt}
\\
$B^+\to K^+\nu\bar\nu$ &\mdfd{$(4.92\pm 0.30)\cdot 10^{-6}$}
&    $ \mdfd{(13\pm 4)}\cdot 10^{-6}$ \cite{Browder:2021hbl,Belle-II:2023esi}
\\
$B^0\to K^{0*}\nu\bar\nu$ & ${(10.13\pm 0.92)}\cdot 10^{-6}$ &
 $\le 1.5\cdot 10^{-5}$ \cite{Grygier:2017tzo}
\\
$B^+\to \tau^+\nu_\tau$ & $(0.88\pm 0.05)\cdot 10^{-4}$ &  $(1.06\pm 0.19)\cdot 10^{-4}$ \cite{Zyla:2020zbs}
\\
$B\to X_s\gamma$ & $(3.46\pm 0.24)\cdot 10^{-4}$ & $(3.32\pm 0.15)\cdot 10^{-4}$
\cite{Zyla:2020zbs}
\\
\hline
$\kpn$ & $(8.60\pm 0.42)\cdot 10^{-11}$ &  $(10.9\pm 3.8)\cdot 10^{-11}$ \cite{NA62:2022hqi} 
\\
 $\klpn$ & $(2.94\pm 0.15)\cdot 10^{-11}$ &   $\le 3.0\cdot 10^{-9}$ \cite{Ahn:2018mvc} 
\\
$(\ksm)_{\rm SD}$& {$(1.85\pm 0.12)\cdot 10^{-13}$} &   $\le 2.1\cdot 10^{-10}$
\cite{LHCb:2020ycd}
\\
$K_L\to\pi^0e^+e^-(+)$ &  $(3.48^{+ 0.92}_{-0.80})\cdot 10^{-11}$      &  $\le 28\cdot 10^{-11}$ \cite{AlaviHarati:2003mr}
\\
$K_L\to\pi^0e^+e^-(-)$ &  $(1.57^{+ 0.61}_{-0.49})\cdot 10^{-11}$      &  
\\
$K_L\to\pi^0\mu^+\mu^-(+)$ &  $(1.39^{+ 0.27}_{-0.25})\cdot 10^{-11}$      &  $\le38\cdot 10^{-11}$ \cite{AlaviHarati:2000hs}
\\
$K_L\to\pi^0\mu^+\mu^-(-)$ &  $(0.95^{+ 0.21}_{-0.20})\cdot 10^{-11}$      &  
\\
\hline
\end{tabular}
}
\renewcommand{\arraystretch}{1.0}
\caption{\label{tab:SMBRBV2}
  \small
  Results for very  rare $B$ and $K$ decay branching ratios using the strategy of   \cite{Buras:2021nns,Buras:2022wpw}. The signs $(\pm)$ in
  $K_L\to\pi^0\ell^+\ell^-$ correspond to the constructive and the destructive intereference between directly and indirectly CP-violating contributions. The result for $B^+\to K^+\nu\bar\nu$ uses most recent formfactors from HPQCD collaboration \cite{Parrott:2022zte,Parrott:2022rgu,Parrott:2022smq}.}
\end{table}

Inserting then experimental values of $\Delta M_q$ into  (\ref{CMFV61}) and
using  the most recent LQCD values of $\hat B_q$ from  \cite{Dowdall:2019bea},
as listed in Table~\ref{tab:input},
one finds the results for $B_{s,d}\to\mu^+\mu^-$ and the remaining  rare $B$ decays in Table~\ref{tab:SMBRBV2}.
Similar, setting the experimental value of $|\varepsilon_K|$ into
  (\ref{R11}) and (\ref{R12a}) and including all theoretical uncertainties and
experimental ones from $|\varepsilon_K|$ and $\beta$ in (\ref{betagamma}) one finds the results for $\kpn$ and $\klpn$ and subsequently for the remaining
rare $K$ decays in Table~\ref{tab:SMBRBV2}.

These are the most precise  SM predictions for decays in question
to date. In particular in the case of $K\to\pi\nu\bar\nu$ they supersede the
widely cited 2015 results   \cite{Buras:2015qea}
\be\label{SM1}
\mathcal{B}(\kpn)_\text{SM}= (8.4\pm 1.0)\times 10^{-11}\,, \quad
\mathcal{B}(\klpn)_\text{SM}=(3.4\pm0.6)\times 10^{-11}, \quad (2015),
\ee
that are clearly out of date as stressed recently in a note by the author \cite{Buras:2022irq}. Using our strategy the uncertainties in the two branching ratios
have been reduced by a factor of $2.4$ and $4.0$, respectively.

Relative to \cite{Buras:2021nns,Buras:2022wpw} the predictions for
$K_L\to\pi^0\ell^+\ell^-$ are new. Moreover, we added to the error in
the prediction for $(\ksm)_{\rm SD}$ the uncertainty from the indirect CP violation pointed out recently in \cite{Brod:2022khx}. Adding it in quadrature the error has
been increased from $5.4\%$ to $6.5\%$. Our final result differs from the one
of these authors because for the CKM parameters they use the UT fit from PDG22 \cite{Workman:2022ynf}
that differs from our strategy. See Section~\ref{CKM} for more details.

Among the results shown in Table~\ref{tab:SMBRBV2} the most interesting
until recently was a $2.7\sigma$ anomaly in $B_s\to\mu^+\mu^-$, but
according to the most recent messages from CMS and HFLAV
this branching ratio has been  increased to $3.45(29)\cdot 10^{-9}$ as given
in Table~\ref{tab:SMBRBV2} thereby eliminating this anomaly. In this context I would like to comment on 
the widely cited by experimentalists SM prediction from \cite{Beneke:2019slt}
$3.66(12)\cdot 10^{-9}$. It is based on NLO QCD \cite{Buchalla:1992zm,Buchalla:1993bv,Misiak:1999yg,Buchalla:1998ba},  NNLO QCD \cite{Hermann:2013kca}, NLO electroweak \cite{Bobeth:2013tba} 
and QED corrections calculated in  \cite{Beneke:2019slt}. However, it does not
properly represent the SM value because the inclusive value of $\vcb$ has been
used to obtain it. As shown in \cite{Buras:2022wpw}, for the exclusive value
of $\vcb$ one finds $3.18(12)\cdot 10^{-9}$. Interestingly the CMS2022 result
alone with  $3.83(42)\cdot 10^{-9}$ agrees perfectly  with our $\vcb$ independent result in Table~\ref{tab:SMBRBV2} which is based on all the perturbative calculations listed above but uses (\ref{CMFV61}) to eliminate $\vcb$. In fact
our SM prediction  has been obtained several months before the new CMS value
\cite{Buras:2022wpw}.

The recent result on
$B^+\to K^+\nu\bar\nu$ from Belle II with \mdfd{$(13\pm4)\times 10^{-6}$} \cite{Belle-II:2023esi}
visibly {\em above} the SM prediction is also interesting 
but the experimental error is still large\footnote{\mdfd{Several recent analyses of $B\to K(K^*)\nu\bar\nu$ decays 
  \cite{Becirevic:2023aov,Bause:2023mfe,Allwicher:2023syp,Dreiner:2023cms} were motivated by this data.}}.
  We are looking forward to the
final CMS and Belle II  analyses and the corresponding ones from LHCb and ATLAS
so that more precise values on both branching ratios will be available from HFLAV.  \mdfd{We should remark that our result for $B^+\to K^+\nu\bar\nu$, similar to Belle II result,   does not include $10\%$ upward shift from a tree-level contribution pointed out in \cite{Kamenik:2009kc}. Otherwise it would be $(5.53\pm 0.30)\times 10^{-6}$ in perfect agreement with the result of HPQCD collaboration
that included this contribution \cite{Parrott:2022zte,Parrott:2022rgu,Parrott:2022smq}.}

In this context a number of important comments should be made.
This method for obtaining
precise SM predictions  has been questioned by a few  flavour researchers
who claim the superiority of global fits in obtaining SM predictions
over the novel methods developed in \cite{Buras:2003td,Buras:2021nns,Buras:2022wpw}
that allowed to remove the sensitivity of SM predictions not only to
$\vcb$ but also to $\gamma$. The criticism is related to the second item in our
proposal, namely the use
of the experimental values for $\Delta F=2$ observables in this strategy, with the goal to obtain SM predictions. The claim is that the presence of
NP in the $\Delta F=2$ observables would invalid the full procedure.

In my view, that is supported by a number of my colleagues, this criticism misses the following important point. The
only assumption made in our procedure is that $\Delta F=2$ observables in (\ref{loop}) are not infected by NP. In a global fit this assumption is made for
many additional observables and the chance of an infection is much larger.
One should also stress that the formulae used to obtain the four ratios in
 (\ref{CMFV61}), (\ref{R11})
and (\ref{R12a}), are only valid
in the SM and in the SM world there are no NP contributions. Therefore, if
one wants to obtain  {{\em genuine}} SM predictions for rare decay branching ratios using these ratios, 
it is simply mandatory to set, in the formulae (\ref{CMFV61}), (\ref{R11})
and (\ref{R12a}),
the quantities in (\ref{loop})
to their experimental values. If one day it will turn out that
      NP infects $\Delta F=2$ processes, then anyway one will have
      to repeat the full analysis in a NP model that will result in 
      predictions for rare decays in this particular model, not
      in the SM. 

      One can also give a simpler argument for the validity of this strategy.
      Formulae (\ref{CMFV61}), (\ref{R11}) and (\ref{R12a}) represent SM
      correlations between chosen $\Delta F=1$ branching ratios and the 
      $\Delta F=2$ observables in question. Setting $\Delta M_s$ to its experimental value gives automatically the SM prediction for $B_s\to\mu^+\mu^-$ and
      similarly for the other three branching ratios. Note that in the case
      of (\ref{R11}) and (\ref{R12a}) these are not just correlations between
      $\kpn$ and $\klpn$ branching ratios and $|\varepsilon_K|$ but with the latter raised to appropiate power so that $\vcb$ and $\gamma$ dependences are eliminated. 
           
      I do hope very much that this underlines again
      the important role of correlations
      between various observables, not only within the SM but also in any model
      as discussed at length in \cite{Buras:2020xsm,Buras:2013ooa,Buras:2015yca}. In my view before doing any global fit it is useful to find first
      these correlations 
      and compare them with data. Within the SM they allow to reduce the
      dependence  on the CKM parameters to the minimum.
\boldmath
\subsection{Rapid Test for the $\Delta F=2$ Sector}\label{rapid}
\unboldmath
Having set the SM expressions for $\Delta F=2$ observables to their
      experimental values we are now in the position to determine the
      CKM parameters. However, before doing it, it is mandatory to perform a
      rapid test to be sure that the resulting CKM parameters are not infected
      by NP. To this end, instead of inserting the formulae in a computer
      program right away it is useful to 
      construct first a $\vcb-\gamma$ plot \cite{Buras:2021nns,Buras:2022wpw} with three bands resulting  separately from $\Delta M_s$, $\Delta M_d$ and
      $|\varepsilon_K|$ constraint and in the latter case imposing the
      constraint from $S_{\psi K_S}$. The superiority of the  $\vcb-\gamma$ plot
      with respect to $\vcb$ and  $\gamma$ over UT plots
      has been recently emphasized in  \cite{Buras:2022nfn}.

\begin{figure}[t!]
  \centering%
  \includegraphics[width=0.70\textwidth]{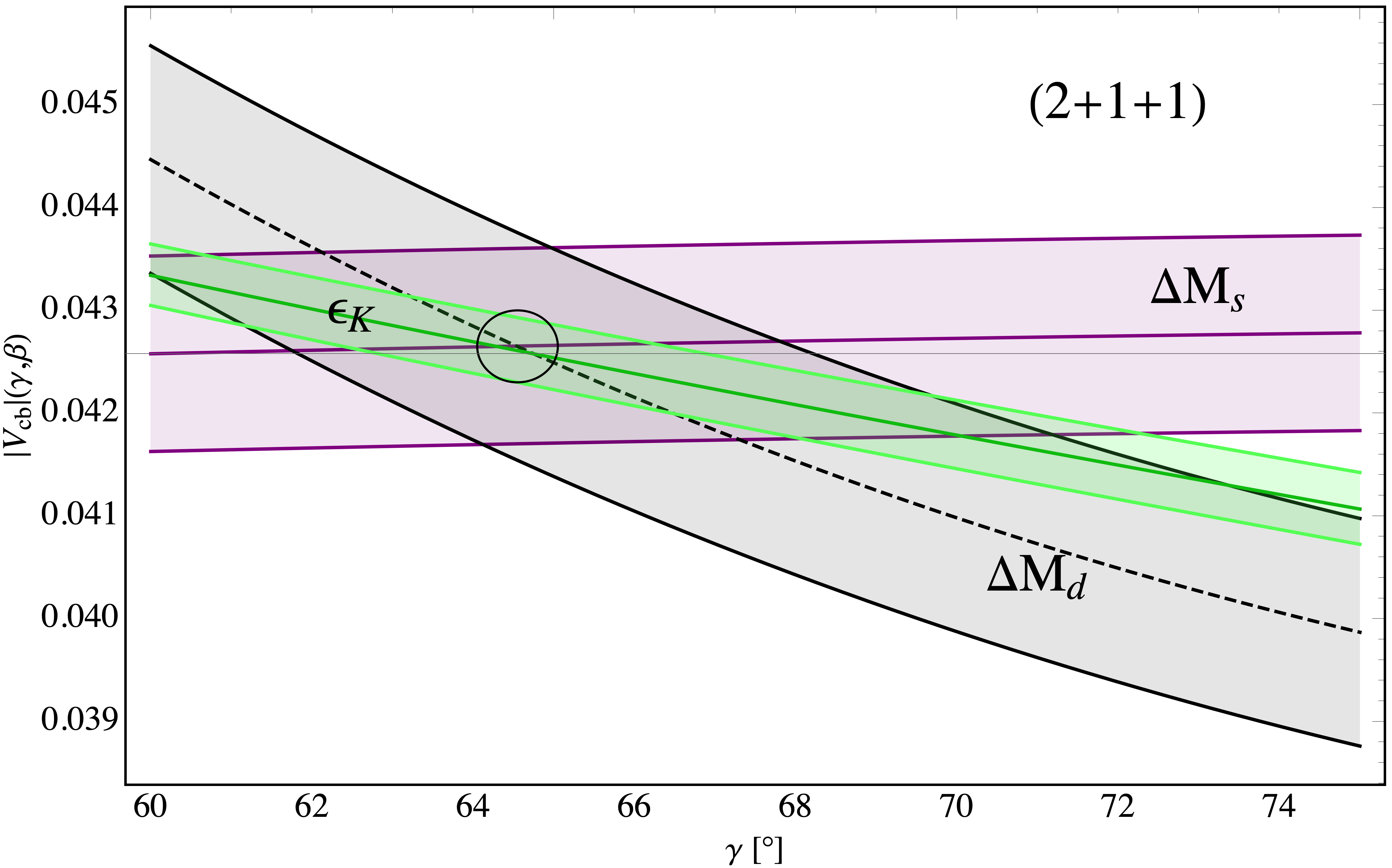}\\
 \includegraphics[width=0.70\textwidth]{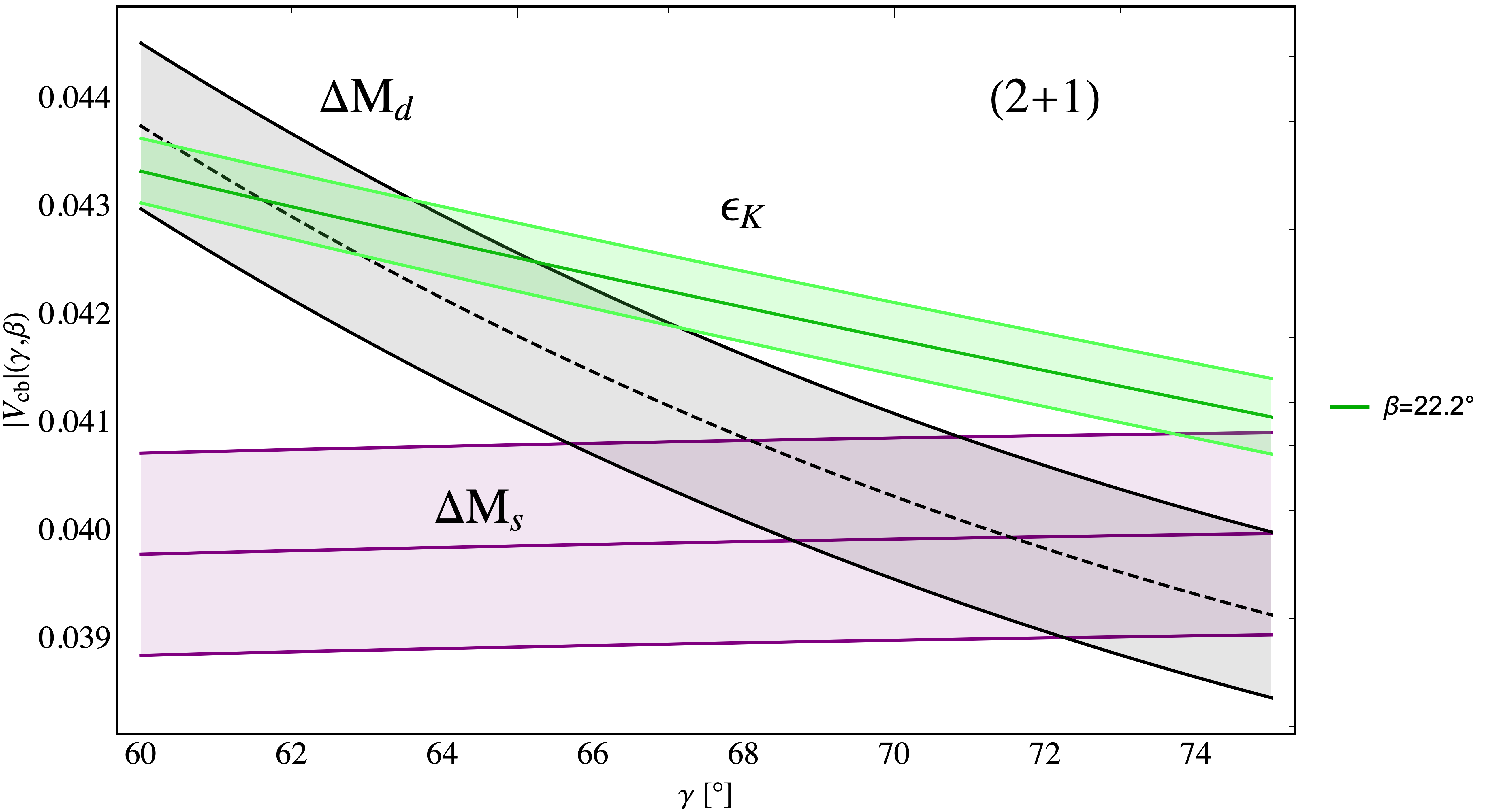}\\
\includegraphics[width=0.70\textwidth]{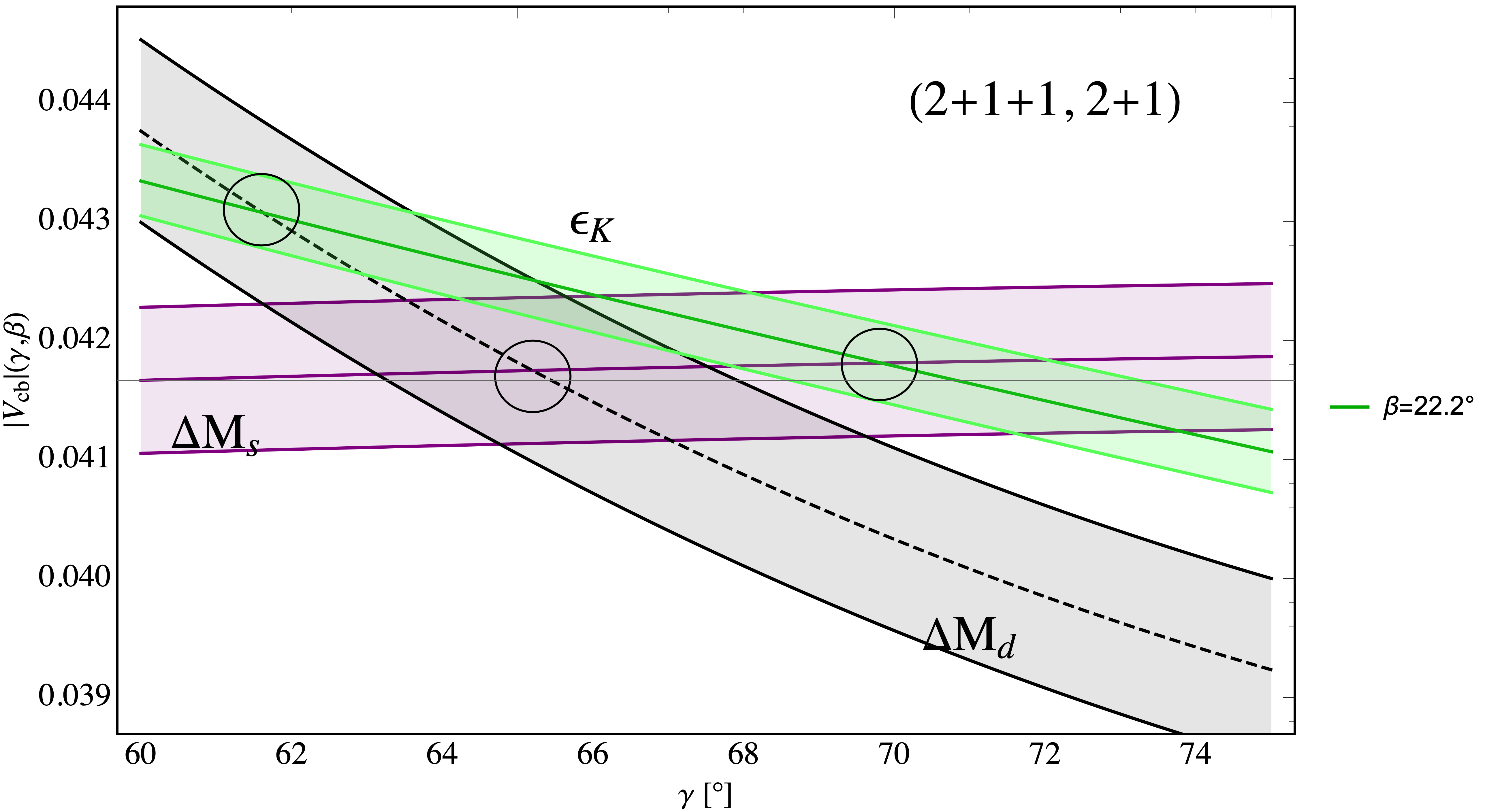}%
\caption{\it {Three rapid tests of NP infection in the $\Delta F=2$ sector taken from \cite{Buras:2022wpw} as explained in the text. The values of $\vcb$ extracted from $\varepsilon_K$, $\Delta M_d$ and  $\Delta M_s$ as functions of $\gamma$. $2+1+1$ flavours (top), $2+1$ flavours (middle), average of $2+1+1$ and $2+1$
    cases (bottom). The green band represents experimental $S_{\psi K_S}$ constraint on $\beta$.}
\label{fig:5}}
\end{figure}

The plots in Fig.~\ref{fig:5}, taken from \cite{Buras:2022wpw}, illustrate
three  {\em rapid tests} of NP infection of the  $\Delta F=2$ sector.
      The test is {\em negative} if these 
      three bands cross each other at a small common area in this plane so
      that unique values of $\vcb$ and $\gamma$ are found. Otherwise it is {\em positive} signalling NP infection. Indeed, 
      as seen in the first $\vcb-\gamma$ plot in  Fig.~\ref{fig:5} that is based on $2+1+1$ LQCD hadronic matrix elements  \cite{Dowdall:2019bea},
      the SM $\vcb-\gamma$ bands resulting from $\varepsilon_K$, $\Delta M_d$ and  $\Delta M_s$ after imposition of the  $S_{\psi K_S}$ constraint, turn out to provide such  unique values of $\vcb$ and $\gamma$. No sign of NP infection in this case. On the other hand, as seen in the remaining
      two plots in   Fig.~\ref{fig:5},  this is not the case if $2+1$
      or the average of $2+1+1$ and $2+1$ hadronic matrix elements LQCD are used. In these two cases the test turns out to be {\em positive}.

 Explicitly these three bands in the $2+1+1$ case are represented by the expressions  \cite{Buras:2022nfn}

 \boldmath
$\underline{|\varepsilon_K|}$
\unboldmath
\bea
\label{epsilon}
\vcb&=&42.6\times 10^{-3}\,\left[\frac{\sin(64.6^\circ)}{\sin\gamma} \right]^{0.491}
\left[\frac{\sin(\beta)}{\sin(22.2^\circ)} \right]^{0.256}
\left[\frac{0.7625}{\hat B_K}\right]^{0.294} 
\left[\frac{|\varepsilon_K|}{2.224\times 10^{-3}} \right]^{0.294}\,,
\eea

\boldmath
$\underline{\Delta M_d}$
\unboldmath
\bea
\label{DMD1}
\vcb&=& 42.6\times 10^{-3}\,\left[\frac{\sin(64.6^\circ)}{\sin\gamma} \right]
\left[\frac{210.6\mev}{\sqrt{\hat B_{B_d}}F_{B_d}}\right]
\left[\frac{2.307}{S_0(x_t)}\right]^{0.5}
\left[\frac{0.5521}{\eta_B}\right]^{0.5}\,
\left[\frac{\Delta M_d}{0.5065/{\rm ps}} \right]^{0.5}\,,
\eea

\boldmath
$\underline{\Delta M_s}$
\unboldmath
\bea
\label{DMS1}
\vcb&=& \left[\frac{41.9\times 10^{-3}}{G(\beta,\gamma)}\right]
\left[\frac{256.1\mev}{\sqrt{\hat B_{B_s}}F_{B_s}}\right]
\left[\frac{2.307}{S_0(x_t)}\right]^{0.5}
\left[\frac{0.5521}{\eta_B}\right]^{0.5}\,
\left[\frac{\Delta M_s}{17.749/{\rm ps}} \right]^{0.5}\,~~~~~~~~
\eea
with $\hat{B}_K =0.7625(97)$~\cite{FlavourLatticeAveragingGroupFLAG:2021npn} and the remaining parameters
given in Table~\ref{tab:input}. Moreover,
\be\label{Gbg}
 G(\beta,\gamma)=
1 +\frac{\lambda^2}{2}(1-2 \sin\gamma\cos\beta)\,.
\ee
Further details on these formulae can be found
in \cite{Buras:2021nns,Buras:2022wpw,Buras:2022nfn}.

Consequently,  with the presently known values of the non-perturbative parameters from LQCD in Table~\ref{tab:input} and the experimental value of $\beta$, the SM is performing in the $\Delta F=2$ sector very well. No NP is required
in this sector to describe the data.
This test will improve with the
  reduction of the uncertainties in $\hat B_K$, $\sqrt{\hat B_{B_d}}F_{B_d}$,
  $\sqrt{\hat B_{B_s}}F_{B_s}$ and $\beta$. Therefore it is very important
  that several LQCD collaborations perform simulations with 2+1+1 flavours.

 All this can also be seen with the help of the following, practically CKM free,
SM relation between the four $\Delta F=2$ observables  in (\ref{loop}) which we present here for the first time. It reads
 \be\label{RNEW}
\boxed{\frac{ |\varepsilon_K|^{1.18}}{\Delta M_d\,\Delta M_s}=(8.22\pm 0.18)\times 10^{-5}\, \left(\frac{\sin \beta}{\sin 22.2^\circ}\right)^{1.027} K\,{\rm ps^2},}
\ee
where
\be
K=\left(\frac{\hat B_K}{0.7625}\right)^{1.18} 
\left[\frac{210.6\mev}{\sqrt{\hat B_{B_d}}F_{B_d}}\right]^2\,
\left[\frac{256.1\mev}{\sqrt{\hat B_{B_s}}F_{B_s}}\right]^2\,=\, 1.00\pm0.07\,.
\ee
Similar to the relations (\ref{R11}) and (\ref{R12a}) the dependence on $\vcb$
drops out and the one on $\gamma$ being  negligible is included in the
uncertainty varying $\gamma$ in the range $60^\circ\le\gamma\le 70^\circ$.
Inserting the experimental values of the three $\Delta F=2$ observables on the l.h.s one finds for this ratio $(8.26\pm0.06)\times 10^{-5}$. Consequently,  with the presently known values of the non-perturbative parameters from LQCD in Table~\ref{tab:input} and the present value of $\beta$ from $S_{\psi K_S}$, the SM is performing in the $\Delta F=2$ sector indeed very well. However with the $2+1$ flavours the central
value on the r.h.s of (\ref{RNEW}) decreases to $(6.29\pm 0.18)\times 10^{-5}$
so that the fact that this ratio  agrees with the data for present values of hadronic parameters with $2+1+1$ flavours and the experimental value of $\beta$ is remarkable.

  {What if the rapid test turns out to be {\em positive} one day. Then it is safer
  to just compare the SM predictions for the ratios of branching ratios like the ones in
  (\ref{CMFV61}), (\ref{R11}) and (\ref{R12a}) which being independent
  of CKM parameters are valid in the SM independently of NP present in $\Delta
  F=2$ processes. In this case the restriction of the fit of the CKM parameters
  to $\Delta F=2$ processes is mainly motivated by the desire to avoid the
  involvement of the tensions between different determinations of $\vcb$ and
  $\vub$. However, with the present accuracy of the hadronic parameters the
  present rapid test is clearly negative.}

  It is possible that one can determine CKM parameters by increasing the
  number of observables beyond $\Delta F=2$ observables used by us, but then
  it should be an obligation to perform a rapid test using $\vcb-\gamma$
  plot that includes additional observables before one could claim
  that the resulting SM predictions for rare branching ratios are indeed
   {{genuine}} SM predictions.

\subsection{CKM parameters}\label{CKM}
      \subsubsection{Our Determination}
The determination of $\gamma$ and $\vcb$
can be further improved by considering first the $\vcb$-independent ratio
$\Delta M_d/\Delta M_s$ from which one derives an accurate formula for $\sin\gamma$
\be
\sin\gamma=\frac{0.983(1)}{\lambda}\sqrt{\frac{m_{B_s}}{m_{B_d}}}\xi
\sqrt{\frac{\Delta M_d}{\Delta M_s}}\,, \qquad \xi=\frac{\sqrt{\hat B_{B_s}}F_{B_s}} {\sqrt{\hat B_{B_d}}F_{B_d}}=1.216(16),
\ee
with the value for $\xi$ from \cite{Dowdall:2019bea}. The advantage of
using this ratio over studying $\Delta M_s$ and $\Delta M_d$ separately
is its $\vcb$-independence and the reduced error on $\xi$ from LQCD relative to the individual errors on
hadronic parameters in $\Delta M_s$ and $\Delta M_d$.

Subsequently
$\vcb$ can be obtained from $\Delta M_s$ that depends only on $\vcb$ and very weakly on $\gamma$ and $\beta$ through $G(\beta,\gamma)$ in (\ref{Gbg}) so that including also $\varepsilon_K$ and $\beta$ in
this analysis 
the following values
of the CKM parameters are found\footnote{$\vus$ is given in Table~\ref{tab:input}.}
\cite{Buras:2022wpw}
\be\label{CKMoutput}
\boxed{\vcb=42.6(4)\times 10^{-3}, \quad 
\gamma=64.6(16)^\circ, \quad \beta=22.2(7)^\circ, \quad \vub=3.72(11)\times 10^{-3}\,}
\ee
and consequently
\be\label{CKMoutput2}
\boxed{\vts=41.9(4)\times 10^{-3}, \qquad \vtd=8.66(14)\times 10^{-3}\,,\qquad
{\IM}\lambda_t=1.43(5)\times 10^{-4}\,,}
\ee
\be\label{CKMoutput3}
\boxed{\bar\varrho=0.164(12),\qquad \bar\eta=0.341(11)\,,}
\ee
where $\lambda_t=V_{ts}^*V_{td}$.

The values of $\vcb$ and $\vub$ are in a very good agreement with the ones obtained
in \cite{Altmannshofer:2021uub} from the $\Delta F=2$ processes alone.
It should be noted that the determination of $\gamma$ in this manner, not
provided in \cite{Altmannshofer:2021uub}, is more
accurate than its present determination from tree-level decays in (\ref{betagamma}). 
This very good agreement between the data and the SM for $\Delta F=2$ observables is an additional strong support for our strategy.
Comparing with the (\ref{alt}) we observe that
the determination of $\vcb$ in the global fit in
\cite{Altmannshofer:2021uub} was indeed infected by NP because  using
the same hadronic input and restricting the analysis to $\Delta F=2$
processes these authors obtained practically the same results for $\vcb$ and $\vub$
as in (\ref{CKMoutput}).

As emphasized in \cite{Buras:2022wpw} and expressed here with the help of
the formulae (\ref{epsilon})-(\ref{DMS1}) and Fig.~\ref{fig:5}, this consistency in the $\Delta F=2$
sector
is only found using
the hadronic matrix elements with $2+1+1$ flavours from the lattice HPQCD collaboration \cite{Dowdall:2019bea}\footnote{Similar results for $\Delta M_d$ and $\Delta M_s$ hadronic
    matrix elements have been obtained within the HQET sum rules in
    \cite{Kirk:2017juj} and \cite{King:2019lal}, respectively.} also used in
\cite{Altmannshofer:2021uub}.
 These values are consistent
 with the inclusive determination of $\vcb$ in \cite{Bordone:2021oof}  and the exclusive ones of $\vub$ from FLAG \cite{FlavourLatticeAveragingGroupFLAG:2021npn}.

 However, let me stress that the values in (\ref{CKMoutput})-(\ref{CKMoutput3}) are only a byproduct of our analysis. Except for $\beta$ obtained using SM expression in (\ref{beta}) I do not have to know other CKM parameters to obtain the SM predictions listed
  in Table~\ref{tab:SMBRBV2}  and in fact to obtain the predictions for all $K$ and $B^0_{d,s}$ branching ratios within the SM. 
  \subsubsection{UTfitter,  CKMfitter and PDG 2022}
  It is instructive to compare our results for the CKM parameters with the most
  recent ones from the UTfitter \cite{UTfit:2022hsi}\footnote{I thank Luca Silvestrini for discussion of these most recent results.}, the CKMfitter and PDG22 \cite{Workman:2022ynf}. These three groups perform  global fits including $\Delta F=2$ observables, tree-level decays relevant for $\vcb$ and $\vub$ determinations and dependently on the analysis some observables like the branching ratio
  for $B_s\to\mu^+\mu^-$ that still could be infected by NP. The same applies to the Cabibbo anomaly which has to be taken somehow into account in a global fit.  The comparison in question
  is made in Table~\ref{tab:comparison}.

  We observe that the   values of $\bar\varrho$, $\bar\eta$, $\vtd$ and $\vub$
  obtained by these three groups are in good agreement with ours, in
  particular the ones from the UTfitter. But the values of $\vcb$ and $\vts$
  are visibly lower with the ones from the UTfitter closer to ours than from the
  CKMfitter and PDG. This in turn implies the SM values of for
  all rare $K$ and $B$ decay branching ratios to be
  lower than ours. For $\kpn$ typically by $(5\pm 1)\%$ dependent
  on the fit. 
  Presently
  these differences do not matter in view of large experimental errors but
  could be relevant in a few years from now.

The main origin
of this difference is the inclusion of the  tree-level determinations of
$\vcb$ for which the tension between exclusive and inclusive determinations
exists. It implies a lower value and larger error on this parameter and consequently when used in the calculations of branching ratios for theoretically clean decays a hadronic pollution of these decays.
In our view the inclusion of the later determinations of $\vcb$ in a global CKM fit or any phenomenological analysis with the goal to predict
SM branching ratios for rare $K$ and $B$ decays is {not a good strategy at present.}
 We think it should be avoided until these tensions are clarified.

 Finally, our value for $\gamma$ is
 closer to its central value from the most recent LHCb measurement in (\ref{betagamma}) with the values from the CKMfitter and PDG by $1.7^\circ$ higher than
 the LHCb value and our only by $0.8^\circ$. It will be interesting to
 make such comparisons when the error on $\gamma$ from LHCb and Belle II will
 go down to $1^\circ$. As the theoretical error for the extraction of $\gamma$ from $B\to DK$ decays is tiny \cite{Brod:2013sga,Backus:2022xhi}, this determination will play
 a very important role for the tests of the SM and also of the $\vcb$ independent correlations between $K$ and $B$ decay branching ratios.

\begin{table}
\centering
\renewcommand{\arraystretch}{1.4}
\resizebox{\columnwidth}{!}{
\begin{tabular}{|l|l|l|l|l|}
\hline
CKM 
& Our Fit & UTfitter
& CKMfitter & PDG22  
\\
\hline \hline
 $\vcb\cdot 10^3$ &  $42.6(4) $  & $42.0(5)$ &    $41.5^{+0.4}_{-0.6}$           & $41.8(8) $
\\
 $\gamma$ &  $ 64.6(16)^\circ$  & $65.1(13)$  &    $65.5(13)^\circ$          & $65.5(15)^\circ$
\\
 $\vub\cdot 10^3$ &  $ 3.72(11) $   & $3.71(9)$   &   $3.67(8)$    &$3.69(11) $
\\
 $\vts\cdot 10^3$ &  $41.9(4) $ & $41.3(5)$  &   $40.7^{+0.4}_{-0.5}$         &$41.1(8) $
\\
$\vtd\cdot 10^3$ & $8.66(14) $ & $8.59(12)$ &   $8.52^{+0.08}_{-0.15}$          & $8.57(20) $
\\
 $\bar\varrho$ &  $0.164(12)$ &$0.162(10)$ &   $0.157^{+0.009}_{-0.005}$          & $0.159(10)$
 \\
 $\bar\eta$ & $0.341(11)$ &$0.347(10)$ &     $0.348^{+0.012}_{-0.005}$           & $0.348(10)$
 \\
\hline
\end{tabular}
}
\renewcommand{\arraystretch}{1.0}
\caption{\label{tab:comparison}
  \small
  Comparison of the CKM output using the  strategy of   \cite{Buras:2021nns,Buras:2022wpw} presented here with UTfitter \cite{UTfit:2022hsi}, CKMfitter and PDG22 \cite{Workman:2022ynf}.}
\end{table}

  \subsection{SM Predictions for $\vcb$-Independent Ratios}\label{ratios}
Among the 16 $\vcb$-independent ratios presented in \cite{Buras:2021nns}
those that correlate $B$ and $K$ branching ratios depend on $\gamma$ and $\beta$. With the results in (\ref{CKMoutput})
at hand we can calculate them. The explicit expressions for these
ratios as functions of $\beta$ and $\gamma$ are given in \cite{Buras:2021nns} and their compact collection can be found in
\cite{Buras:2022nrb}. Here we just list the final results using (\ref{CKMoutput}) which were not given there. Moreover in the case of the ratios $R_5$ and
$R_7$ we use the most recent results for the formfactors entering $\mathcal{B}(B^+\to K^+\nu\bar\nu)$ from the HPQCD collaboration \cite{Parrott:2022zte,Parrott:2022rgu,Parrott:2022smq}. 
\be
\boxed{R_0(\beta)=\frac{\mathcal{B}(\kpn)}{\mathcal{B}(\klpn)^{0.7}}
={(2.03\pm 0.11)}\times 10^{-3}\,,
}
\label{eq:R0}
\ee
    \be\label{SR1}
\boxed{R_{\rm SL}=\frac{\BR(\ksm)_{\rm SD}}{\BR(\klpn)}=(6.29\pm0.52)\times10^{-3}\,,}
\ee

\be\label{R1}
\boxed{R_1(\beta,\gamma)=\frac{\mathcal{B}(\kpn)}{\left[{\overline{\mathcal{B}}}(B_s\to\mu^+\mu^-)\right]^{1.4}}= 53.69\pm2.75\,,}
  \ee
  \be \label{R2}
  \boxed{ R_2(\beta,\gamma)=\frac{\mathcal{B}(\kpn)}{\left[{\mathcal{B}}(B_d\to\mu^+\mu^-)\right]^{1.4}}= (8.51\pm0.47)\times 10^{-3}\,,}
   \ee

   \be\label{R3}
   \boxed{R_3(\beta,\gamma)=\frac{\mathcal{B}(\klpn)}{\left[{\overline{\mathcal{B}}}(B_s\to\mu^+\mu^-)\right]^{2}}= (2.08\pm0.16)\times 10^6\,.}
   \ee
   \be\label{R4}
\boxed{R_4(\beta,\gamma)=\frac{\mathcal{B}(\klpn)}{\left[{\mathcal{B}}(B_d\to\mu^+\mu^-)\right]^{2}}=(2.90\pm0.24)\times 10^9\,,}
   \ee

\be\label{R5}
\boxed{R_5(\beta,\gamma)=\frac{\mathcal{B}(\kpn)}{\left[\mathcal{B}(B^+\to K^+\nu\bar\nu)\right]^{1.4}}=\mdfd{(2.32\pm0.14)\times 10^{-3}\,,}}
\ee
\be\label{R6}
\boxed{R_6(\beta,\gamma)=\frac{\mathcal{B}(\kpn)}{\left[\mathcal{B}(B^0\to K^{0*}\nu\bar\nu)\right]^{1.4}}=(8.82\pm1.21)\times 10^{-4}\,.}
\ee

\be\label{R7}
\boxed{R_7=\frac{\mathcal{B}(B^+\to K^+\nu\bar\nu)}{{\overline{\mathcal{B}}}(B_s\to\mu^+\mu^-)}=\mdfd{(1.30\pm0.07)\times 10^{3}\,.}}
\ee

\be\label{R8}
\boxed{R_8=\frac{\mathcal{B}(B^0\to K^{*0}\nu\bar\nu)}{{\overline{\mathcal{B}}}(B_s\to\mu^+\mu^-)}=(2.62\pm0.25)\times 10^{3}\,.}
    \ee
    One can check that the uncertainties in the ratios above are smaller than
    the ones one would find by calculating them by means of the results in
    Table~\ref{tab:SMBRBV2} because some uncertainties cancel in the ratio when
    they are calculated directly using the expressions in
\cite{Buras:2021nns,Buras:2022wpw}.
         
    The ratios $R_{9}$ and $R_{10}$ involve only  $|\varepsilon_K|$ and $\Delta M_{s,d}$ which were used in the rapid test and in the determination of the CKM parameters from (\ref{loop}) so that we can skip them here. Presently, most interesting are the ratios in
 (\ref{CMFV61}), (\ref{R11}) and (\ref{R12a} for which we find
 \be\label{R13}
  \boxed{R_s=\frac{\mathcal{B}(B_s\to\mu^+\mu^-)}{\Delta M_s}=
    (2.13\pm0.07)\times 10^{-10}\text{ps}\,,}
\ee
   \be\label{R14}
\boxed{R_d=\frac{\mathcal{B}(B_d\to\mu^+\mu^-)}{\Delta M_d}= (2.02\pm0.08)\times 10^{-10}\text{ps}\,,}
\ee
  
  \be\label{R11b}
  \boxed{R_{11}(\beta,\gamma)=\frac{\mathcal{B}(\kpn)}{|\varepsilon_K|^{0.82}}=
      (1.31\pm0.06)\times 10^{-8}\,,    }
  \ee
  \be\label{R12ab}
  \boxed{R_{12}(\beta,\gamma)=\frac{\mathcal{B}(\klpn)}{|\varepsilon_K|^{1.18}}=
 (3.87\pm0.13)\times 10^{-8}\,.
  }
  \ee

\boldmath
\section{SM Predictions for $H_1\to H_2\mu^+\mu^-$ Branching Ratios}\label{sec:3b}
\unboldmath
The semi-leptonic transitions $b\to s\ell^+\ell^-$ have been left out in
\cite{Buras:2021nns,Buras:2022wpw} because of larger hadronic uncertainties than
is the case of decays listed in Table~\ref{tab:SMBRBV2}. However, in fact
having the result for $\vts$ in (\ref{CKMoutput2}) we can next calculate all branching ratios
involved in the $B$-physics anomalies. To this end we use a very useful
formula \cite{Altmannshofer:2021uub}
\be\label{BALT}
\mathcal{B}(H_1\to H_2\mu^+\mu^-)_\text{SM}^{[q^2_{\text{min}},q^2_{\text{max}}]}=\vts^2
a_{H_1\to H_2}^{[q^2_{\text{min}},q^2_{\text{max}}]},
\ee
where the superscript $[q^2_{\text{min}},q^2_{\text{max}}]$ indicates $q^2$ bin.
For each decay mode the authors of  \cite{Altmannshofer:2021uub} calculated
the numerical coefficients in front of $\vts^2$ for one broad $q^2$ bin below the narrow charmonium resonances and one broad bin above. For the numerical coefficients in (\ref{BALT})
they find \cite{Altmannshofer:2021uub}
\be\label{ALT1}
a_{B^+\to K^+}^{[1.1,6]}=(1.00\pm0.16)\times 10^{-4}, \qquad 
a_{B^+\to K^+}^{[15,22]}=(0.61\pm0.06)\times 10^{-4},
\ee
\be
a_{B^0\to K^{*0}}^{[1.1,6]}=(1.36\pm0.16)\times 10^{-4}, \qquad 
a_{B^0\to K^{*0}}^{[15,19]}=(1.39\pm0.15)\times 10^{-4},
  \ee
\be
a_{B_s\to \phi}^{[1.1,6]}=(1.54\pm0.14)\times 10^{-4}, \qquad 
a_{B_s\to \phi}^{[15,19]}=(1.30\pm0.12)\times 10^{-4}\,.
  \ee
\be
a_{\Lambda_b\to \Lambda}^{[1.1,6]}=(0.30\pm0.16)\times 10^{-4}, \qquad 
  a_{\Lambda_b\to \Lambda}^{[15,20]}=(2.07\pm0.21)\times 10^{-4}.
\ee

These results are based on \cite{Straub:2018kue,Straub:2015ica,Gubernari:2018wyi}. However, recently new results from HPQCD collaboration with $2+1+1$ flavours
\cite{Parrott:2022zte,Parrott:2022rgu,Parrott:2022smq} for $B^+\to K^+$ formfactors became
available from which we extract
\be
a_{B^+\to K^+}^{[1.1,6]}={(1.09\pm0.07)\times 10^{-4}, \quad 
a_{B^+\to K^+}^{[15,22]}=(0.66\pm0.05)\times 10^{-4}, }\quad (\text{HPQCD22}).
\ee
We will use these results instead of (\ref{ALT1}) in what follows.

Using then these coefficients together with $\vts$ in (\ref{CKMoutput2}) we obtain
the results for various branching ratios  listed in Table~\ref{tab:SMBRBV1}.
We compare them with the data and list the pulls in the last column. 
While some pulls are in the ballpark of $(2-3)\sigma$, we find 
a $-4.8\sigma$ anomaly in $B_s\to\phi\mu^+\mu^-$ in the lower $q^2$ bin. This
finding agrees with the one of \cite{Altmannshofer:2021uub}. Similarly a large 
pull of {$-4.2\sigma$} in the low $q^2$ bin in  $B^+\to K^+\mu^+\mu^-$ has  been found recently by HPQCD
colaboration \cite{Parrott:2022rgu}. With our CKM parameters it is further
increased to {$-4.4\sigma$}\footnote{We thank Will Parrott from the HPQCD collaboration for confirming this result.}.
These appear to be the largest anomalies in  single branching ratios.

It should be noted that for all branching ratios in  Table~\ref{tab:SMBRBV1}
one can construct, with the help of $\Delta M_s$, the CKM independent ratios
as in the previous section. Here we just present the results for
the two among them in  the low $q^2$ bin that exhibit  the largest pulls mentioned above.
We find
\be\label{R13new}
  \boxed{R_{13}=\frac{\mathcal{B}(B^+\to K^+\mu^+\mu^-)}{\Delta M_s}=
   {(1.076\pm0.073)\times 10^{-8} }   \left[\frac{256.1\mev}{\sqrt{\hat B_{B_s}}F_{B_s}}\right]^2\,\text{ps}\,,\qquad [1.1,6]   }
  \ee
and
  \be\label{R14new}
  \boxed{R_{14}=\frac{\mathcal{B}(B^+\to \phi\mu^+\mu^-)}{\Delta M_s}=
    (1.523\pm0.138)\times 10^{-8} \left[\frac{256.1\mev}{\sqrt{\hat B_{B_s}}F_{B_s}}\right]^2\,\text{ps}\,,\qquad [1.1,6] .}
\ee
Including the uncertainty in $\sqrt{\hat B_{B_s}}F_{B_s}$ we find
\be\label{R13newf}
  \boxed{R_{13}=
    {(1.076\pm0.087)\times 10^{-8}\,\text{ps}},\qquad R_{13}^{\text{EXP}}=
    (0.668\pm0.038)\times 10^{-8}\,\text{ps}\,,\qquad [1.1,6]\,,}
  \ee
\be\label{R14newf}
  \boxed{R_{14}=
    (1.523\pm0.154)\times 10^{-8}\,\text{ps},\qquad R_{14}^{\text{EXP}}=
    (0.794\pm0.056)\times 10^{-8}\,\text{ps}\,,\qquad [1.1,6]}
  \ee
and the pulls
{$-4.3\sigma$} and $-4.5\sigma$, respectively. The reduction of the pulls relative
to the ones for branching ratios in Table~\ref{tab:SMBRBV1} 
originates in the larger error from the hadronic uncertainty in 
$\sqrt{\hat B_{B_s}}F_{B_s}$ than the uncertainty in $\vts$ obtained from
the $\Delta F=2$ fit that involves also $\Delta M_d$, $|\varepsilon_K|$ and
$S_{\psi K_S}$. But the advantage over the branching ratios themselves is that
these ratios are free from any CKM dependence.

Importantly, the experimental branching ratios are for most of the branching
ratios in Table~\ref{tab:SMBRBV1} below the SM predictions which expresses
the anomalies widely discussed in the literature.
It should also be emphasized that studying various differential distributions,
various asymmeteries $S_i$ and $A_i$ as proposed in \cite{Altmannshofer:2008dz}
or $P_i(P_i^\prime)$ variables proposed in \cite{Descotes-Genon:2013vna} that suffer from smaller
hadronic  and parameteric uncertainties than branching ratios themselves the pulls in $B\to K(K^*)\mu^+\mu^-$ could turn out to be larger. Yet, just testing the
branching ratios themselves is much simpler and can give already some indications
on the presence of NP.

\boldmath
\section{SM Predictions for $b\to s\nu\bar\nu$ Transitions}\label{sec:3c}
\unboldmath
Several SM branching ratios for $B$ decays with neutrino pair in the final
state beyond those discussed by us above have been calculated in  \cite{Bause:2021cna} with a much lower value
of $\vts=39.7\times 10^{-3}$ than used by us\footnote{For $B_s\to\phi\nu\bar\nu$ ref. \cite{Li:2022xxs} confirms the results of \cite{Bause:2021cna} using practically the same value of $\vts$.}. We present in Table~\ref{tab:nunubar} the
corresponding results with our value of $\vts$ in (\ref{CKMoutput2}). They
are typicaly by $11\%$ higher than the ones in \cite{Bause:2021cna}.
The interest in the $B$  decays with neutrino pair in the final  was already significant for years\footnote{See \cite{Altmannshofer:2009ma,Buras:2014fpa} and the references therein.} but it increased recently
due to the BELLE II experiment \cite{Kou:2018nap} as seen in
\cite{Bause:2021cna,Felkl:2021uxi,Descotes-Genon:2020buf,He:2021yoz,Browder:2021hbl,Descotes-Genon:2021doz,Descotes-Genon:2022gcp} and \mdfd{most recently in
\cite{Becirevic:2023aov,Bause:2023mfe,Allwicher:2023syp,Dreiner:2023cms} 
  after   the BELLE II result in \cite{Belle-II:2023esi}. }

\begin{table}
\centering
\renewcommand{\arraystretch}{1.4}
\resizebox{\columnwidth}{!}{
\begin{tabular}{|ll|l|l|l|}
\hline
Decay & $[q^2_{\text{min}},q^2_{\text{max}}]$
& Branching Ratio (SM)
& Branching Ratio (EXP) & Pull
\\
\hline \hline
$B^+\to K^+\mu^+\mu^-$ & $[1.1,6]$ &
{$(1.91\pm0.15)\cdot 10^{-7}$ }     &    $(1.186\pm0.068)\cdot 10^{-7}$ \cite{LHCb:2014cxe} & {$-4.40$}
\\
$B^+\to K^+\mu^+\mu^-$ & $[15,22]$ &
{$(1.16\pm0.09)\cdot 10^{-7}$ }    &    $(0.847\pm0.050)\cdot 10^{-7}$ \cite{LHCb:2014cxe}  & {$-3.04$}
\\
$B_d^0\to K^{*0}\mu^+\mu^-$ & $[1.1,6]$ &
$(2.39\pm0.28)\cdot 10^{-7}$    &    $(1.68\pm0.15)\cdot 10^{-7}$ \cite{LHCb:2016ykl} & $-2.23$
\\
$B_d^0\to K^{*0}\mu^+\mu^-$ & $[15,19]$ &
$(2.44\pm0.26)\cdot 10^{-7}$    &    $(1.74\pm0.14)\cdot 10^{-7}$  \cite{LHCb:2016ykl} & $-2.37$
\\
$B_s\to \phi\mu^+\mu^-$ & $[1.1,6]$ &
$(2.70\pm0.25)\cdot 10^{-7}$    &    $(1.41\pm0.10)\cdot 10^{-7}$ \cite{LHCb:2021zwz} & $-4.80$
\\
$B_s\to \phi\mu^+\mu^-$ & $[15,19]$ &
$(2.28\pm0.21)\cdot 10^{-7}$     &    $(1.85\pm0.13)\cdot 10^{-7}$ \cite{LHCb:2021zwz} & $-1.74$
\\
$\Lambda_b\to \Lambda\mu^+\mu^-$ & $[1.1,6]$ &
$(0.53\pm0.28)\cdot 10^{-7}$    &    $(0.44\pm0.31)\cdot 10^{-7}$ \cite{LHCb:2015tgy}& $-0.21$
\\
$\Lambda_b\to \Lambda\mu^+\mu^-$ & $[15,20]$ &
$(3.63\pm0.37)\cdot 10^{-7}$    &    $(6.00\pm1.34)\cdot 10^{-7}$ \cite{LHCb:2015tgy} & $+1.70$
\\
\hline
\end{tabular}
}
\renewcommand{\arraystretch}{1.0}
\caption{\label{tab:SMBRBV1}
  \small
  SM predictions for  $H_1\to H_2\mu^+\mu^-$ branching ratios with
  $[q^2_{\text{min}},q^2_{\text{max}}]$ compared with the data. Last column
  gives the pull.}
\end{table}

\begin{table}
\centering
\renewcommand{\arraystretch}{1.4}
\resizebox{\columnwidth}{!}{
\begin{tabular}{|l|l|l|l|}
\hline
Decay 
& SM1 & SM2
& Data
\\
\hline \hline
 $B_s\to\phi\nu\bar\nu$ &  $(10.9\pm0.7)\cdot 10^{-6}$  &  $(10.9\pm0.9)\cdot 10^{-6}$   &  $\le 5.4\cdot 10^{-3}$ \cite{DELPHI:1996ohp}
\\
 $B_s\to K^0\nu\bar\nu$ &  $ (1.4\pm 0.3)\ \cdot 10^{-7}$  & $ (1.4\pm 0.3)\ \cdot 10^{-7}$    & 
\\
 $B_s\to K^{0*}\nu\bar\nu$ &  $ (4.0\pm 0.3)\ \cdot 10^{-7}$   & $ (4.0\pm 0.4)\ \cdot 10^{-7}$  & 
\\
 $B^0_d\to X_S\nu\bar\nu$ &  $(3.1\pm0.3)\cdot 10^{-5}$ &  $(3.1\pm0.4)\cdot 10^{-5}$     &  $\le 6.4\cdot 10^{-4}$ \cite{ALEPH:2000vvi}
\\
 $B^+\to X_s\nu\bar\nu$ &  $(3.3\pm0.3)\cdot 10^{-5}$&  $(3.3\pm0.4)\cdot 10^{-5}$    &  $\le 6.4\cdot 10^{-4}$ \cite{ALEPH:2000vvi}
\\
\hline
\end{tabular}
}
\renewcommand{\arraystretch}{1.0}
\caption{\label{tab:nunubar}
  \small
  Selective results for SM branching ratios using the strategy of   \cite{Buras:2021nns,Buras:2022wpw} obtained by using the results in \cite{Bause:2021cna}.
  SM1: with our value of $\vts$ in  (\ref{CKMoutput2}), SM2: removal of $\vts$
  using $\Delta M_s$.}
\end{table}

\boldmath
\section{Direct Route to SM Predictions for $H_1\to H_2\mu^+\mu^-$ Branching Ratios and $b\to s\nu\bar\nu$}\label{sec:3d}
\unboldmath
It should be stressed that the predictions in Sections~\ref{sec:3b}
  and ~\ref{sec:3c} go beyond the main strategy of removing CKM parameters
    from the analyses and we report here how our results in the previous two sections would change if we eliminated  $\vts$ with the help of $\Delta M_s$
    and setting its value to the experimental one. This procedure
    is a bit safer as the results are  expected to be
       more stable under future
      modifications of $\vts$ due to possible changes in
      non-perturbative parameters in the $\Delta F=2$ system beyond those
      relevant for $\Delta M_s$. Basically the present uncertainty from $\vts^2$
      of $1.9\%$ obtained from the full $\Delta F=2$ fit increases to $4.4\%$. But as the uncertainties in the formfactors have presently  a significantly larger impact on the error in the final preditions these changes are small. In particular the central values are not modified because,  as seen in Fig.~\ref{fig:5}, $\Delta M_s$ being only very
      weakly dependent on $\gamma$  plays an important role in the determination of $\vcb$ in the full $\Delta F=2$ fit.
      We just quote a few examples in the modifications of the resulting errors:
      \be
      B^+\to K^+\mu^+\mu^-~ ([1.1,6]): \quad {(1.91\pm0.15)\cdot 10^{-7}\rightarrow
      (1.91\pm0.17)\cdot 10^{-7}\,,}
      \ee
      \be
B^0\to K^{*0}\mu^+\mu^-~([1.1,6]): \quad  (2.39\pm0.28)\cdot 10^{-7}\rightarrow 
(2.39\pm0.30)\cdot 10^{-7}\,,
\ee
\be
B_d^0\to K^{*0}\mu^+\mu^-~ ([15,19]): \quad (2.44\pm0.26)\cdot 10^{-7}\rightarrow
(2.44\pm0.28)\cdot 10^{-7}\,,
\ee
\be
B_s\to \phi\mu^+\mu^-~([1.1,6]):\quad (2.70\pm0.25)\cdot 10^{-7}\rightarrow
(2.70\pm0.27)\cdot 10^{-7}\,,
\ee
\be
B_s\to \phi\mu^+\mu^-~([15,19]): \quad 
(2.28\pm0.21)\cdot 10^{-7}\rightarrow  (2.28\pm0.23)\cdot 10^{-7}\,.
\ee

In the case of final states with $\nu\bar\nu$ these changes are described in
Table~\ref{tab:nunubar}.

\section{Exclusive and Hybrid Scenarios}\label{EXHYB}
But what if one day experts agree on the basis of tree-level decays that
the values of the CKM parameters differ from those that are listed  in (\ref{CKMoutput}). For instance one could consider, as done in  \cite{Buras:2022wpw},
the following two well defined scenarios based on tree-level decays. First the
EXCLUSIVE one
\be\label{FLAGVUB1}
\vcb=39.21(62)\times 10^{-3},\qquad 
\vub=3.61(13)\times 10^{-3}, \qquad \qquad ({\rm EXCLUSIVE})
\ee
that summarize preliminary results from FLAG2022 and 
the HYBRID one
in which the value for $\vcb$ is the inclusive one from \cite{Bordone:2021oof} and the exclusive
one for $\vub$ as above:
\be\label{HYBRID}
{\vcb=42.16(50)\times 10^{-3},\qquad 
\vub=3.61(13)\times 10^{-3}, \qquad {(\rm HYBRID)}.}
\ee

The important point to be stressed here is the following one. The SM
predictions for those $\vcb$ independent ratios,
defined in \cite{Buras:2021nns} and evaluated in  Section~\ref{ratios} 
that  are independent of all CKM parameters,  
will be modified in the future only by changes in hadronic parameters.
In the ratios involving $K$ decays the value of $\beta$ matters and could
modify the ratios in addition in the future. However, as seen in (\ref{R11}) and (\ref{R12a}), for $R_{11}$ and $R_{12}$  the  $\gamma$ dependence is negligible.
 Other ratios can depend significantly on $\gamma$ and $\beta$ and this dependence is exhibited in numerous plots in \cite{Buras:2021nns}.

But the values of the branching ratios and also of $\Delta M_s$, $\Delta M_d$ and
$\varepsilon_K$ will change, in particular by much in the exclusive scenario.
However, it will happen in a correlated manner with correlations simply described by the $\vcb$-independent ratios.

In particular, as analysed in detail in \cite{Buras:2022wpw}, in the exclusive
scenario significant anomalies in $\Delta M_s$, $\Delta M_d$ and $\varepsilon_K$ will be found, while several ones in $B$ decays will be removed or decreased.
For instance all branching ratios in Tables~\ref{tab:SMBRBV1} and \ref{tab:nunubar} will be suppressed by a factor $0.847$ reducing significantly the present anomalies
and in the case of the $B_s\to\mu^+\mu^-$ decay removing it completely. But the room
for NP opened in the $\Delta F=2$ sector will significantly weaken
the constraints on NP from this sector. As seen in \cite{Buras:2022wpw}, in the hybrid scenario the results
do not differ by much from the ones presented here but have larger errors
dominantly due to larger error on $\gamma$ than in (\ref{CKMoutput}).

   \section{Searching for Footprints of NP Beyond the SM}\label{sec:5a}
 Having the results from our strategy at hand, 
 the simplest route to find out whether there is some NP, once the experimental values of many branching ratios will be known, is in my view the following one:

 {\bf Step 1:}

 Comparison of CKM-independent ratios like (\ref{CMFV61}) with experiment.
 In the case of $R_s$ there was already a sign of NP.
The SM prediction for $R_s$ and the resulting SM prediction
for $B_s\to\mu^+\mu^-$ branching ratio differed by $2.7\sigma$ from the data.
However, this difference has been reduced by much  due to the recent CMS result.
Once $B_d\to\mu^+\mu^-$ branching ratio will be measured, similar test
will be possible for $R_d$ and other decays like $B\to K(K^*)\nu\bar\nu$.
Even
more interesting are the pulls in the low $q^2$ bin in the ratios
$R_{13}$ and $R_{14}$ involving 
 $B^+\to K^+\mu^+\mu^-$ {($-4.3\sigma$)} and $B_s\to \phi\mu^+\mu^-$ ($-4.5\sigma$), respectively.

When the branching ratios for $\kpn$, $\klpn$ and other rare $K$ decays will be measured, SM predictions will be tested through ratios like $R_{11}$ and
$R_{12}$ that depend practically only on $\beta$.

It should be stressed that all these ratios do not involve the assumption
of the absence of NP in $\Delta F=2$ observables and in the case of the
sign of NP in the ratio it could come from the $\Delta F=1$ observable or $\Delta F=2$
observable or even both.

{\bf Step 2:}

Once the rapid test in Section~\ref{rapid} is found to be negative one can set the
$\Delta F=2$ observables to their experimental values. This allows
to predict the branching ratios either by means of the $\vcb$-independent
ratios or just using the CKM parameters determined exclusively from
$\Delta F=2$ observables. The results for the branching ratios are collected in 
Tables~\ref{tab:SMBRBV2}, \ref{tab:SMBRBV1} and \ref{tab:nunubar}.
Similarly, one can calculate those $\vcb$-independent ratios of \cite{Buras:2021nns} that depend on $\beta$ and $\gamma$. The results are given in
(\ref{eq:R0})-(\ref{R12ab}) and  (\ref{R13new})-(\ref{R14newf}).

Following these steps, future measurements of all branching ratios calculated in the present paper will hopefully tell us  what is the pattern of deviations from their SM predictions allowing us to select some favourite BSM models.
Indeed in this context various $\vcb$ independent ratios of branching ratios considered by us, both
independent  of $\beta$ and $\gamma$ and dependent on them and calculated by us
in Section~\ref{ratios} will provide a good test of the SM.
Similarly 
$\vcb-\gamma$ plots \cite{Buras:2021nns,Buras:2022wpw,Buras:2022nfn} 
will play an important role, in particular if $\beta$ and $\gamma$ will
be determined in tree-level non-leptonic $B$ decays that are likely to
receive only very small  NP contributions. However, this may still take some time. Then also the comparison with
the values in (\ref{CKMoutput}) will be possible. Moreover, beyond the SM
the ratios $R_i$ will depend on $\vcb$ so that its value will be necessary
for the study of NP contributions. Therefore, it is very important that this
direct route to $\vcb$ through trevel decays is continued with all technology
we have to our disposal.

\section{Conclusions and Outlook}\label{sec:4}
We have pointed out that the most straightforward method for obtaining
SM predictions for rare $K$ and $B$ decays is to study those SM correlations between the branching ratios and $\Delta F=2$ observables that do not depend
or depend minimally on the CKM parameters. The standard method is to determine
the latter first through global fits and subsequently insert the resulting
values into SM formulae. In view of the mounting evidence for NP in
semi-leptonic $B$ decays the resulting values of the CKM parameters are likely to be infected by NP if such decays are included in a global fit. Inserting them
in the SM expressions for rare decays in question will obviously not provide 
{genuine} SM predictions for their branching ratios.

The determination of the CKM parameters {\em exclusively} from tree-level decays
could in principle reduce the dependence of CKM parameters on NP\footnote{Nonetheless, NP can also affect these decays as stressed in \cite{Brod:2014bfa,Lenz:2019lvd,Iguro:2020ndk,Cai:2021mlt,Bordone:2021cca}.} and the prospects of their determination in the coming years are good \cite{Lenz:2022kal}.
However,
the present tensions between inclusive and exclusive determination of $\vcb$
is a stumbling block on this route to SM predictions of branching ratios that are very sensitive to $\vcb$ \cite{Buras:2021nns}. As demonstrated
in \cite{Buras:2022wpw} going this route using the exclusive determination of $\vcb$ would result in very different predictions than obtained by using
the corresponding inclusive route. The recent analysis in  \cite{DeBruyn:2022zhw} demonstrates this problem as well.

As proposed very recently in \cite{Faroughy:2022dyq} the sum  $\vtd^2+\vts^2$ could also
be accessed through CKM suppressed top decays at the LHC. We note that this
would provide another route to $\vcb$ through
\be
\vtd^2+\vts^2=\vcb^2[G^2(\beta,\gamma)+\lambda^2 R_t^2],\qquad R_t=\frac{\sin\gamma}{\sin(\beta+\gamma)}\,,
\ee
where $G(\beta,\gamma)$ is given in (\ref{Gbg}) with $\beta$ and $\gamma$ determined
through tree-level non-leptonic $B$ decays. This would avoid the use of
presently controversial value of $\vcb$ from tree-level semi-leptonic $B$ decays. This would also provide another test of  our values of the CKM parameters.
Using them we find
\be
\vtd^2+\vts^2=42.8(4)\times 10^{-3}\,.
\ee

It should be emphasized that to obtain precise SM predictions like the
  ones in Table~\ref{tab:SMBRBV2} it is crucial to choose the proper
  pairs of observables. For instance combining $\kpn$ with $\Delta M_s$ or
  $B_s\to\mu^+\mu^-$ with $\varepsilon_K$ would not allow us precise predictions
  for $\kpn$ and $B_s\to\mu^+\mu^-$ even after the elimination of the $\vcb$ because of the left-over $\gamma$ dependence in both cases. Moreover
  selecting a subset of optimal observables for a given SM prediction with the
  goal of removing the CKM dependence   avoids the assumption of the absence of NP in other observables that enter necessarily  a global fit.

  It is known from numerous studies that NP could have significant impact
  on $\Delta F=2$ observables, in particular in the presence of left-right
  operators which have enhanced hadronic matrix elements and 
  their contributions to $\Delta F=2$ processes are additionally enhanced
  through QCD renormalization group effects. One could then ask the question
  how in the presence of significant NP contributions to semi-leptonic decays
  one could avoid large contributions to $\Delta F=2$ observables. Some
  answers are given in the 4321  model \cite{DiLuzio:2017vat,DiLuzio:2018zxy}
  and in a number of analyses by Isidori's group \cite{Cornella:2019hct,Fuentes-Martin:2019ign,Fuentes-Martin:2020luw,Fuentes-Martin:2020hvc,Crosas:2022quq}
in which a specific
  flavour structure allows to suppress the
  contributions to $\Delta F=2$
  processes from the leptoquark $U_1$, heavy $Z^\prime$, $G^\prime$ and vector-like
  fermions while allowing for their sizeable contributions to semileptonic decays. Yet, the fact that the SM performs so well in the $\Delta F=2$ sector when
  the HPQCD results \cite{Dowdall:2019bea} are used  puts even stronger
  constraints on  NP model constructions than in the past. Therefore it is crucial that other LQCD collaborations perform $2+1+1$ calculations of $\Delta F=2$ hadronic matrix
  elements.

  In the spirit of the last word in the title of our paper it will be of
  interest to see one day whether the archipelago of $\Delta F=2$ observables
  will be as little infected by NP as has been the Galapagos archipelago by Covid-19
  and other pandemics in the past. The expressions in Section~\ref{rapid}
  provide
  a {\em rapid test} in this context. This test will improve with the
  reduction of the uncertainties in $\hat B_K$, $\sqrt{\hat B_{B_d}}F_{B_d}$,
    $\sqrt{\hat B_{B_s}}F_{B_s}$ and $\beta$.

  However, even if this test would fail and  NP
  would infect $\Delta F=2$ observables,  the $\vcb$ independent ratios
  introduced in \cite{Buras:2003td,Buras:2021nns}, in particular those free
  of the CKM parameters, will offer  excellent tests of the SM dynamics. Such tests will be truly powerful
  when the uncertainties on $\gamma$ and $\beta$ from tree-level decays
  will be reduced in the coming years.

  We are looking forward to the days  on which numerous results
  presented in Tables~\ref{tab:SMBRBV2}- \ref{tab:nunubar},  in the 
  formulae (\ref{eq:R0})-(\ref{R12ab}) and (\ref{R13new})-(\ref{R14newf}) will be compared with improved experimental data.
  In particular it is of great interest to see whether 
the anomalies in the low $q^2$ bin in
 $B^+\to K^+\mu^+\mu^-$ {($4.4\sigma$)} and $B_s\to \phi\mu^+\mu^-$ ($4.8\sigma$)
will remain even if the violation of the lepton flavour universality  in
semi-leptonic decays would disappear.

 {\bf Acknowledgements}
 I would like to thank Christine Davies and Will Parrott for the discussions
 on their recent paper \cite{Parrott:2022rgu}.
 Many thanks go also to Elena Venturini for the most enjoyable collaboration that was influential for the results presented here. The comments of 
 Fulvia de Fazio and Luca Silvestrini on the V1 of our paper are
 highly appreciated as well as discussions with Jason Aebischer, Marcela Bona, Paolo Gambino, Andreas Kronfeld, Jacky Kumar  and Laura Reina. Financial support from the Excellence Cluster ORIGINS,
funded by the Deutsche Forschungsgemeinschaft (DFG, German Research
Foundation), 
Excellence Strategy, EXC-2094, 390783311 is acknowledged.

\appendix

\begin{table}[!tb]
\center{\begin{tabular}{|l|l|}
\hline
$m_{B_s} = 5366.8(2)\mev$\hfill\cite{Zyla:2020zbs}	&  $m_{B_d}=5279.58(17)\mev$\hfill\cite{Zyla:2020zbs}\\
$\Delta M_s = 17.749(20) \,\text{ps}^{-1}$\hfill \cite{Zyla:2020zbs}	&  $\Delta M_d = 0.5065(19) \,\text{ps}^{-1}$\hfill \cite{Zyla:2020zbs}\\
{$\Delta M_K = 0.005292(9) \,\text{ps}^{-1}$}\hfill \cite{Zyla:2020zbs}	&  {$m_{K^0}=497.61(1)\mev$}\hfill \cite{Zyla:2020zbs}\\
$S_{\psi K_S}= 0.699(17)$\hfill\cite{Zyla:2020zbs}
		&  {$F_K=155.7(3)\mev$\hfill  \cite{FlavourLatticeAveragingGroupFLAG:2021npn}}\\
	$|V_{us}|=0.2253(8)$\hfill\cite{Zyla:2020zbs} &
 $|\eps_K|= 2.228(11)\cdot 10^{-3}$\hfill\cite{Zyla:2020zbs}\\
$F_{B_s}$ = $230.3(1.3)\mev$ \hfill \cite{FlavourLatticeAveragingGroupFLAG:2021npn} & $F_{B_d}$ = $190.0(1.3)\mev$ \hfill \cite{Aoki:2021kgd}  \\
$F_{B_s} \sqrt{\hat B_s}=256.1(5.7) \mev$\hfill  \cite{Dowdall:2019bea}&
$F_{B_d} \sqrt{\hat B_d}=210.6(5.5) \mev$\hfill  \cite{Dowdall:2019bea}
\\
 $\hat B_s=1.232(53)$\hfill\cite{Dowdall:2019bea}        &
 $\hat B_d=1.222(61)$ \hfill\cite{Dowdall:2019bea}          
\\
{$m_t(m_t)=162.83(67)\GeV$\hfill\cite{Brod:2021hsj} }  & {$m_c(m_c)=1.279(13)\GeV$} \\
{$S_{tt}(x_t)=2.303$} & {$S_{ut}(x_c,x_t)=-1.983\times 10^{-3}$} \\
    $\eta_{tt}=0.55(2)$\hfill\cite{Brod:2019rzc} & $\eta_{ut}= 0.402(5)$\hfill\cite{Brod:2019rzc}\\
$\kappa_\varepsilon = 0.94(2)$\hfill \cite{Buras:2010pza}	&
$\eta_B=0.55(1)$\hfill\cite{Buras:1990fn,Urban:1997gw}\\
$\tau_{B_s}= 1.515(4)\,\text{ps}$\hfill\cite{Amhis:2016xyh} & $\tau_{B_d}= 1.519(4)\,\text{ps}$\hfill\cite{Amhis:2016xyh}   
\\	       
\hline
\end{tabular}  }
\caption {\textit{Values of the experimental and theoretical
    quantities used as input parameters. For future 
updates see FLAG  \cite{FlavourLatticeAveragingGroupFLAG:2021npn}, PDG \cite{Zyla:2020zbs}  and HFLAV  \cite{Aoki:2019cca,HFLAV:2022pwe}. 
}}
\label{tab:input}
\end{table}

\renewcommand{\refname}{R\lowercase{eferences}}

\addcontentsline{toc}{section}{References}

\bibliographystyle{JHEP}

\small

\bibliography{Bookallrefs}

\providecommand{\href}[2]{#2}\begingroup\raggedright\begin{thebibliography}{100}

\bibitem{Cerri:2018ypt}
A.~Cerri, V.~V. Gligorov, S.~Malvezzi, J.~Martin~Camalich, and J.~Zupan, {\it
  {Opportunities in Flavour Physics at the HL-LHC and HE-LHC}},
  \href{http://arxiv.org/abs/1812.07638}{{\tt arXiv:1812.07638}}.

\bibitem{Bediaga:2018lhg}
{\bf LHCb} Collaboration, R.~Aaij et~al., {\it {Physics case for an LHCb
  Upgrade II - Opportunities in flavour physics, and beyond, in the HL-LHC
  era}},  \href{http://arxiv.org/abs/1808.08865}{{\tt arXiv:1808.08865}}.

\bibitem{NA62:2022nah}
{\it {Searches for new physics with high-intensity kaon beams}},
  \href{http://arxiv.org/abs/2204.13394}{{\tt arXiv:2204.13394}}.

\bibitem{FlavourLatticeAveragingGroupFLAG:2021npn}
{\bf Flavour Lattice Averaging Group (FLAG)} Collaboration, Y.~Aoki et~al.,
  {\it {FLAG Review 2021}},  {\em Eur. Phys. J. C} {\bf 82} (2022), no.~10 869,
  [\href{http://arxiv.org/abs/2111.09849}{{\tt arXiv:2111.09849}}].

\bibitem{Blum:2022wsz}
T.~Blum et~al., {\it {Discovering new physics in rare kaon decays}},  3, 2022.
\newblock \href{http://arxiv.org/abs/2203.10998}{{\tt arXiv:2203.10998}}.

\bibitem{Kirk:2017juj}
M.~Kirk, A.~Lenz, and T.~Rauh, {\it {Dimension-six matrix elements for meson
  mixing and lifetimes from sum rules}},  {\em JHEP} {\bf 12} (2017) 068,
  [\href{http://arxiv.org/abs/1711.02100}{{\tt arXiv:1711.02100}}]. [Erratum:
  JHEP 06, 162 (2020)].

\bibitem{Cirigliano:2011ny}
V.~Cirigliano, G.~Ecker, H.~Neufeld, A.~Pich, and J.~Portoles, {\it {Kaon
  Decays in the Standard Model}},  {\em Rev.~Mod.~Phys.} {\bf 84} (2012) 399,
  [\href{http://arxiv.org/abs/1107.6001}{{\tt arXiv:1107.6001}}].

\bibitem{Buchalla:1995vs}
G.~Buchalla, A.~J. Buras, and M.~E. Lautenbacher, {\it {Weak decays beyond
  leading logarithms}},  {\em Rev.~Mod.~Phys.} {\bf 68} (1996) 1125--1144,
  [\href{http://arxiv.org/abs/hep-ph/9512380}{{\tt hep-ph/9512380}}].

\bibitem{Buras:2011we}
A.~J. Buras, {\it {Climbing NLO and NNLO Summits of Weak Decays: 1988-2023}},
  {\em Phys. Rept.} {\bf 1025} (2023) 1--64,
  [\href{http://arxiv.org/abs/1102.5650}{{\tt arXiv:1102.5650}}].

\bibitem{Buras:2020xsm}
A.~J. Buras, {\em {Gauge Theory of Weak Decays}}.
\newblock Cambridge University Press, 6, 2020.

\bibitem{Aebischer:2022vky}
J.~Aebischer, A.~J. Buras, and J.~Kumar, {\it {On the Importance of Rare Kaon
  Decays: A Snowmass 2021 White Paper}},  3, 2022.
\newblock \href{http://arxiv.org/abs/2203.09524}{{\tt arXiv:2203.09524}}.

\bibitem{Cabibbo:1963yz}
N.~Cabibbo, {\it {Unitary Symmetry and Leptonic Decays}},  {\em Phys. Rev.
  Lett.} {\bf 10} (1963) 531--533. [648(1963)].

\bibitem{Kobayashi:1973fv}
M.~Kobayashi and T.~Maskawa, {\it {CP Violation in the Renormalizable Theory of
  Weak Interaction}},  {\em Prog.~Theor.~Phys.} {\bf 49} (1973) 652--657.

\bibitem{Altmannshofer:2021uub}
W.~Altmannshofer and N.~Lewis, {\it {Loop-induced determinations of $V_{ub}$
  and $V_{cb}$}},  {\em Phys. Rev. D} {\bf 105} (2022), no.~3 033004,
  [\href{http://arxiv.org/abs/2112.03437}{{\tt arXiv:2112.03437}}].

\bibitem{Buras:1990fn}
A.~J. Buras, M.~Jamin, and P.~H. Weisz, {\it {Leading and next-to-leading QCD
  corrections to $\varepsilon$ parameter and $B^0-\bar{B}^0$ mixing in the
  presence of a heavy top quark}},  {\em Nucl.~Phys.} {\bf B347} (1990)
  491--536.

\bibitem{Herrlich:1993yv}
S.~Herrlich and U.~Nierste, {\it {Enhancement of the $K_L - K_S$ mass
  difference by short distance QCD corrections beyond leading logarithms}},
  {\em Nucl.~Phys.} {\bf B419} (1994) 292--322,
  [\href{http://arxiv.org/abs/hep-ph/9310311}{{\tt hep-ph/9310311}}].

\bibitem{Herrlich:1995hh}
S.~Herrlich and U.~Nierste, {\it {Indirect CP violation in the neutral kaon
  system beyond leading logarithms}},  {\em Phys.~Rev.} {\bf D52} (1995)
  6505--6518, [\href{http://arxiv.org/abs/hep-ph/9507262}{{\tt
  hep-ph/9507262}}].

\bibitem{Herrlich:1996vf}
S.~Herrlich and U.~Nierste, {\it {The Complete $|\Delta S|=2$ Hamiltonian in
  the Next-To-Leading Order}},  {\em Nucl.~Phys.} {\bf B476} (1996) 27--88,
  [\href{http://arxiv.org/abs/hep-ph/9604330}{{\tt hep-ph/9604330}}].

\bibitem{Brod:2011ty}
J.~Brod and M.~Gorbahn, {\it {Next-to-Next-to-Leading-Order Charm-Quark
  Contribution to the CP Violation Parameter $\varepsilon_K$ and $\Delta
  M_K$}},  {\em Phys.~Rev.~Lett.} {\bf 108} (2012) 121801,
  [\href{http://arxiv.org/abs/1108.2036}{{\tt arXiv:1108.2036}}].

\bibitem{Brod:2010mj}
J.~Brod and M.~Gorbahn, {\it {$\epsilon_K$ at Next-to-Next-to-Leading Order:
  The Charm-Top-Quark Contribution}},  {\em Phys.~Rev.} {\bf D82} (2010)
  094026, [\href{http://arxiv.org/abs/1007.0684}{{\tt arXiv:1007.0684}}].

\bibitem{Brod:2019rzc}
J.~Brod, M.~Gorbahn, and E.~Stamou, {\it {Standard-Model Prediction of
  $\epsilon_K$ with Manifest Quark-Mixing Unitarity}},  {\em Phys. Rev. Lett.}
  {\bf 125} (2020), no.~17 171803, [\href{http://arxiv.org/abs/1911.06822}{{\tt
  arXiv:1911.06822}}].

\bibitem{Brod:2021qvc}
J.~Brod, S.~Kvedarait\.{e}, and Z.~Polonsky, {\it {Two-loop electroweak
  corrections to the Top-Quark Contribution to \ensuremath{\epsilon}$_{K}$}},
  {\em JHEP} {\bf 12} (2021) 198, [\href{http://arxiv.org/abs/2108.00017}{{\tt
  arXiv:2108.00017}}].

\bibitem{Brod:2022har}
J.~Brod, S.~Kvedaraite, Z.~Polonsky, and A.~Youssef, {\it {Electroweak
  corrections to the Charm-Top-Quark Contribution to $\epsilon_K$}},  {\em
  JHEP} {\bf 12} (2022) 014, [\href{http://arxiv.org/abs/2207.07669}{{\tt
  arXiv:2207.07669}}].

\bibitem{Charles:2004jd}
{\bf CKMfitter Group} Collaboration, J.~Charles et~al., {\it {CP violation and
  the CKM matrix: Assessing the impact of the asymmetric $B$ factories}},  {\em
  Eur.~Phys.~J.} {\bf C41} (2005) 1--131,
  [\href{http://arxiv.org/abs/hep-ph/0406184}{{\tt hep-ph/0406184}}].

\bibitem{Bona:2007vi}
{\bf UTfit} Collaboration, M.~Bona et~al., {\it {Model-independent constraints
  on $\Delta$F=2 operators and the scale of new physics}},  {\em JHEP} {\bf
  0803} (2008) 049, [\href{http://arxiv.org/abs/0707.0636}{{\tt
  arXiv:0707.0636}}].

\bibitem{Straub:2018kue}
D.~M. Straub, {\it {flavio: a Python package for flavour and precision
  phenomenology in the Standard Model and beyond}},
  \href{http://arxiv.org/abs/1810.08132}{{\tt arXiv:1810.08132}}.

\bibitem{DeBlas:2019ehy}
J.~De~Blas et~al., {\it {$\texttt{HEPfit}$: a code for the combination of
  indirect and direct constraints on high energy physics models}},  {\em Eur.
  Phys. J. C} {\bf 80} (2020), no.~5 456,
  [\href{http://arxiv.org/abs/1910.14012}{{\tt arXiv:1910.14012}}].

\bibitem{Buras:2022irq}
A.~J. Buras, {\it {On the Standard Model Predictions for Rare K and B Decay
  Branching Ratios: 2022}},  \href{http://arxiv.org/abs/2205.01118}{{\tt
  arXiv:2205.01118}}.

\bibitem{DeBruyn:2022zhw}
K.~De~Bruyn, R.~Fleischer, E.~Malami, and P.~van Vliet, {\it {New physics in
  $B_{q}^{0}$ mixing: present challenges, prospects, and implications for}},
  {\em J. Phys. G} {\bf 50} (2023), no.~4 045003,
  [\href{http://arxiv.org/abs/2208.14910}{{\tt arXiv:2208.14910}}].

\bibitem{Buras:2021nns}
A.~J. Buras and E.~Venturini, {\it {Searching for New Physics in Rare $K$ and
  $B$ Decays without $|V_{cb}|$ and $|V_{ub}|$ Uncertainties}},  {\em Acta
  Phys. Polon. B} {\bf 53} (9, 2021) A1,
  [\href{http://arxiv.org/abs/2109.11032}{{\tt arXiv:2109.11032}}].

\bibitem{Buras:2022wpw}
A.~J. Buras and E.~Venturini, {\it {The exclusive vision of rare K and B decays
  and of the quark mixing in the standard model}},  {\em Eur. Phys. J. C} {\bf
  82} (2022), no.~7 615, [\href{http://arxiv.org/abs/2203.11960}{{\tt
  arXiv:2203.11960}}].

\bibitem{Bordone:2021oof}
M.~Bordone, B.~Capdevila, and P.~Gambino, {\it {Three loop calculations and
  inclusive $\vcb$}},  {\em Phys. Lett. B} {\bf 822} (2021) 136679,
  [\href{http://arxiv.org/abs/2107.00604}{{\tt arXiv:2107.00604}}].

\bibitem{Alguero:2022wkd}
M.~Alguer\'o, J.~Matias, B.~Capdevila, and A.~Crivellin, {\it {Disentangling
  lepton flavor universal and lepton flavor universality violating effects in
  $b\to s l^+ l^-$ transitions}},  {\em Phys. Rev. D} {\bf 105} (2022), no.~11
  113007, [\href{http://arxiv.org/abs/2205.15212}{{\tt arXiv:2205.15212}}].

\bibitem{Buras:2015qea}
A.~J. Buras, D.~Buttazzo, J.~Girrbach-Noe, and R.~Knegjens, {\it {$ {K}^{+}\to
  {\pi}^{+}\nu \overline{\nu} $ and $ {K}_L\to {\pi}^0\nu \overline{\nu} $ in
  the Standard Model: status and perspectives}},  {\em JHEP} {\bf 11} (2015)
  033, [\href{http://arxiv.org/abs/1503.02693}{{\tt arXiv:1503.02693}}].

\bibitem{Colangelo:2016ymy}
P.~Colangelo and F.~De~Fazio, {\it {Tension in the inclusive versus exclusive
  determinations of $|V_{cb}|$: a possible role of new physics}},  {\em Phys.
  Rev.} {\bf D95} (2017), no.~1 011701,
  [\href{http://arxiv.org/abs/1611.07387}{{\tt arXiv:1611.07387}}].

\bibitem{Buras:2003td}
A.~J. Buras, {\it {Relations between $\Delta M_{s,d}$ and $B_{s,d} \to \mu^+
  \mu^-$ in models with minimal flavour violation}},  {\em Phys.~Lett.} {\bf
  B566} (2003) 115--119, [\href{http://arxiv.org/abs/hep-ph/0303060}{{\tt
  hep-ph/0303060}}].

\bibitem{Crivellin:2022qcj}
A.~Crivellin and J.~Matias, {\it {Beyond the Standard Model with Lepton Flavor
  Universality Violation}},  in {\em {1st Pan-African Astro-Particle and
  Collider Physics Workshop}}, 4, 2022.
\newblock \href{http://arxiv.org/abs/2204.12175}{{\tt arXiv:2204.12175}}.

\bibitem{Dowdall:2019bea}
R.~J. Dowdall, C.~T.~H. Davies, R.~R. Horgan, G.~P. Lepage, C.~J. Monahan,
  J.~Shigemitsu, and M.~Wingate, {\it {Neutral $B$-meson mixing from full
  lattice QCD at the physical point}},  {\em Phys. Rev. D} {\bf 100} (2019),
  no.~9 094508, [\href{http://arxiv.org/abs/1907.01025}{{\tt
  arXiv:1907.01025}}].

\bibitem{deBruyn:2012wk}
K.~De~Bruyn, R.~Fleischer, R.~Knegjens, P.~Koppenburg, M.~Merk, et~al., {\it
  {Probing New Physics via the $B^0_s\to \mu^+\mu^-$ Effective Lifetime}},
  {\em Phys.~Rev.~Lett.} {\bf 109} (2012) 041801,
  [\href{http://arxiv.org/abs/1204.1737}{{\tt arXiv:1204.1737}}].

\bibitem{LHCb:2021dcr}
{\bf LHCb} Collaboration, R.~Aaij et~al., {\it {Simultaneous determination of
  CKM angle $\gamma$ and charm mixing parameters}},  {\em JHEP} {\bf 12} (2021)
  141, [\href{http://arxiv.org/abs/2110.02350}{{\tt arXiv:2110.02350}}].

\bibitem{LHCb:2021awg}
{\bf LHCb} Collaboration, R.~Aaij et~al., {\it {Measurement of the
  $B^0_s\to\mu^+\mu^-$ decay properties and search for the $B^0\to\mu^+\mu^-$
  and $B^0_s\to\mu^+\mu^-\gamma$ decays}},  {\em Phys. Rev. D} {\bf 105}
  (2022), no.~1 012010, [\href{http://arxiv.org/abs/2108.09283}{{\tt
  arXiv:2108.09283}}].

\bibitem{CMS:2020rox}
{\bf CMS} Collaboration, {\it {Combination of the ATLAS, CMS and LHCb results
  on the $B^0_{(s)} \to \mu^+\mu^-$ decays}},
  \href{http://arxiv.org/abs/CMS-PAS-BPH-20-003}{{\tt CMS-PAS-BPH-20-003}}.

\bibitem{ATLAS:2020acx}
{\bf ATLAS} Collaboration, {\it {Combination of the ATLAS, CMS and LHCb results
  on the $B^0_{(s)}\to\mu^+\mu^-$ decays.}},
  \href{http://arxiv.org/abs/ATLAS-CONF-2020-049}{{\tt ATLAS-CONF-2020-049}}.

\bibitem{HFLAV:2022pwe}
{\bf HFLAV} Collaboration, Y.~Amhis et~al., {\it {Averages of $b$-hadron,
  $c$-hadron, and $\tau$-lepton properties as of 2021}},
  \href{http://arxiv.org/abs/2206.07501}{{\tt arXiv:2206.07501}}.

\bibitem{Aaij:2017xqt}
{\bf LHCb} Collaboration, R.~Aaij et~al., {\it {Search for the decays
  $B_s^0\to\tau^+\tau^-$ and $B^0\to\tau^+\tau^-$}},  {\em Phys. Rev. Lett.}
  {\bf 118} (2017), no.~25 251802, [\href{http://arxiv.org/abs/1703.02508}{{\tt
  arXiv:1703.02508}}].

\bibitem{Browder:2021hbl}
T.~E. Browder, N.~G. Deshpande, R.~Mandal, and R.~Sinha, {\it {Impact of $B\to
  K\nu\bar\nu$ measurements on beyond the Standard Model theories}},  {\em
  Phys. Rev. D} {\bf 104} (2021), no.~5 053007,
  [\href{http://arxiv.org/abs/2107.01080}{{\tt arXiv:2107.01080}}].

\bibitem{Belle-II:2023esi}
{\bf Belle-II} Collaboration, I.~Adachi et~al., {\it {Evidence for $B^{+}\to
  K^{+}\nu\bar{\nu}$ Decays}},  \href{http://arxiv.org/abs/2311.14647}{{\tt
  arXiv:2311.14647}}.

\bibitem{Grygier:2017tzo}
{\bf Belle} Collaboration, J.~Grygier et~al., {\it {Search for $B\to
  h\nu\bar{\nu}$ decays with semileptonic tagging at Belle}},  {\em Phys. Rev.}
  {\bf D96} (2017), no.~9 091101, [\href{http://arxiv.org/abs/1702.03224}{{\tt
  arXiv:1702.03224}}]. [Addendum: Phys. Rev.D97,no.9,099902(2018)].

\bibitem{Zyla:2020zbs}
{\bf Particle Data Group} Collaboration, P.~A. Zyla et~al., {\it {Review of
  Particle Physics}},  {\em PTEP} {\bf 2020} (2020), no.~8 083C01.

\bibitem{NA62:2022hqi}
{\bf NA62} Collaboration, M.~Zamkovsk\'y et~al., {\it {Measurement of the very
  rare $K^+ \to \pi^+ \nu \bar\nu$ decay}},  {\em PoS} {\bf DISCRETE2020-2021}
  (2022) 070.

\bibitem{Ahn:2018mvc}
{\bf KOTO} Collaboration, J.~Ahn et~al., {\it {Search for the $K_L \!\to\!
  \pi^0 \nu \overline{\nu}$ and $K_L \!\to\! \pi^0 X^0$ decays at the J-PARC
  KOTO experiment}},  {\em Phys. Rev. Lett.} {\bf 122} (2019), no.~2 021802,
  [\href{http://arxiv.org/abs/1810.09655}{{\tt arXiv:1810.09655}}].

\bibitem{LHCb:2020ycd}
{\bf LHCb} Collaboration, R.~Aaij et~al., {\it {Constraints on the $K^0_S
  \rightarrow \mu^+ \mu^-$ Branching Fraction}},  {\em Phys. Rev. Lett.} {\bf
  125} (2020), no.~23 231801, [\href{http://arxiv.org/abs/2001.10354}{{\tt
  arXiv:2001.10354}}].

\bibitem{AlaviHarati:2003mr}
{\bf KTeV} Collaboration, A.~Alavi-Harati et~al., {\it {Search for the rare
  decay $K_L\to\pi^0 e^+ e^-$}},  {\em Phys. Rev. Lett.} {\bf 93} (2004)
  021805, [\href{http://arxiv.org/abs/hep-ex/0309072}{{\tt hep-ex/0309072}}].

\bibitem{AlaviHarati:2000hs}
{\bf KTEV} Collaboration, A.~Alavi-Harati et~al., {\it {Search for the Decay
  $K_L \to \pi^0 \mu^+ \mu^-$}},  {\em Phys. Rev. Lett.} {\bf 84} (2000)
  5279--5282, [\href{http://arxiv.org/abs/hep-ex/0001006}{{\tt
  hep-ex/0001006}}].

\bibitem{Parrott:2022zte}
{\bf HPQCD} Collaboration, W.~G. Parrott, C.~Bouchard, and C.~T.~H. Davies,
  {\it {Standard Model predictions for $B\to K\ell^+\ell^-$, $B\to K\ell_1^-
  \ell_2^+$ and $B\to K\nu\bar{\nu}$ using form factors from $N_f=2+1+1$
  lattice QCD}},  {\em Phys. Rev. D} {\bf 107} (2023), no.~1 014511,
  [\href{http://arxiv.org/abs/2207.13371}{{\tt arXiv:2207.13371}}].

\bibitem{Parrott:2022rgu}
{\bf HPQCD} Collaboration, W.~G. Parrott, C.~Bouchard, and C.~T.~H. Davies,
  {\it {B\textrightarrow{}K and D\textrightarrow{}K form factors from fully
  relativistic lattice QCD}},  {\em Phys. Rev. D} {\bf 107} (2023), no.~1
  014510, [\href{http://arxiv.org/abs/2207.12468}{{\tt arXiv:2207.12468}}].

\bibitem{Parrott:2022smq}
W.~G. Parrott, C.~Bouchard, and C.~T.~H. Davies, {\it {The search for new
  physics in $B \to K \ell^+\ell^-$ and $B \to K \nu\bar{\nu}$ using precise
  lattice QCD form factors}},  in {\em {39th International Symposium on Lattice
  Field Theory}}, 10, 2022.
\newblock \href{http://arxiv.org/abs/2210.10898}{{\tt arXiv:2210.10898}}.

\bibitem{Brod:2022khx}
J.~Brod and E.~Stamou, {\it {Impact of indirect CP violation on
  $Br(K_S\to\mu^+\mu^-)_{l=0}$}},  {\em JHEP} {\bf 05} (2023) 155,
  [\href{http://arxiv.org/abs/2209.07445}{{\tt arXiv:2209.07445}}].

\bibitem{Workman:2022ynf}
{\bf Particle Data Group} Collaboration, R.~L. Workman, {\it {Review of
  Particle Physics}},  {\em PTEP} {\bf 2022} (2022) 083C01.

\bibitem{Beneke:2019slt}
M.~Beneke, C.~Bobeth, and R.~Szafron, {\it {Power-enhanced leading-logarithmic
  QED corrections to $B_q \to \mu^+\mu^-$}},  {\em JHEP} {\bf 10} (2019) 232,
  [\href{http://arxiv.org/abs/1908.07011}{{\tt arXiv:1908.07011}}].

\bibitem{Buchalla:1992zm}
G.~Buchalla and A.~J. Buras, {\it {QCD corrections to the $\bar s d Z$ vertex
  for arbitrary top quark mass}},  {\em Nucl.~Phys.} {\bf B398} (1993)
  285--300.

\bibitem{Buchalla:1993bv}
G.~Buchalla and A.~J. Buras, {\it {QCD corrections to rare $K$ and $B$ decays
  for arbitrary top quark mass}},  {\em Nucl.~Phys.} {\bf B400} (1993)
  225--239.

\bibitem{Misiak:1999yg}
M.~Misiak and J.~Urban, {\it {QCD corrections to FCNC decays mediated by Z
  penguins and W boxes}},  {\em Phys.~Lett.} {\bf B451} (1999) 161--169,
  [\href{http://arxiv.org/abs/hep-ph/9901278}{{\tt hep-ph/9901278}}].

\bibitem{Buchalla:1998ba}
G.~Buchalla and A.~J. Buras, {\it {The rare decays $K\to\pi \nu\bar\nu$, $B\to
  X \nu\bar\nu$ and $B\to \ell^+\ell^-$: An Update}},  {\em Nucl.~Phys.} {\bf
  B548} (1999) 309--327, [\href{http://arxiv.org/abs/hep-ph/9901288}{{\tt
  hep-ph/9901288}}].

\bibitem{Hermann:2013kca}
T.~Hermann, M.~Misiak, and M.~Steinhauser, {\it {Three-loop QCD corrections to
  $B_s \to \mu^+ \mu^-$}},  {\em JHEP} {\bf 1312} (2013) 097,
  [\href{http://arxiv.org/abs/1311.1347}{{\tt arXiv:1311.1347}}].

\bibitem{Bobeth:2013tba}
C.~Bobeth, M.~Gorbahn, and E.~Stamou, {\it {Electroweak Corrections to $B_{s,d}
  \to \ell^+ \ell^-$}},  {\em Phys.~Rev.} {\bf D89} (2014) 034023,
  [\href{http://arxiv.org/abs/1311.1348}{{\tt arXiv:1311.1348}}].

\bibitem{Becirevic:2023aov}
D.~Be\v{c}irevi\'c, G.~Piazza, and O.~Sumensari, {\it {Revisiting $B\rightarrow
  K^{(*)} \nu {\bar{\nu }}$ decays in the Standard Model and beyond}},  {\em
  Eur. Phys. J. C} {\bf 83} (2023), no.~3 252,
  [\href{http://arxiv.org/abs/2301.06990}{{\tt arXiv:2301.06990}}].

\bibitem{Bause:2023mfe}
R.~Bause, H.~Gisbert, and G.~Hiller, {\it {Implications of an enhanced $B \to K
  \nu \bar \nu$ branching ratio}},  \href{http://arxiv.org/abs/2309.00075}{{\tt
  arXiv:2309.00075}}.

\bibitem{Allwicher:2023syp}
L.~Allwicher, D.~Becirevic, G.~Piazza, S.~Rosauro-Alcaraz, and O.~Sumensari,
  {\it {Understanding the first measurement of $\mathcal{B}(B\to K \nu
  \bar{\nu})$}},  \href{http://arxiv.org/abs/2309.02246}{{\tt
  arXiv:2309.02246}}.

\bibitem{Dreiner:2023cms}
H.~K. Dreiner, J.~Y. G\"unther, and Z.~S. Wang, {\it {The Decay $B\to
  K\nu\bar{\nu}$ at Belle II and a Massless Bino in R-parity-violating
  Supersymmetry}},  \href{http://arxiv.org/abs/2309.03727}{{\tt
  arXiv:2309.03727}}.

\bibitem{Kamenik:2009kc}
J.~F. Kamenik and C.~Smith, {\it {Tree-level contributions to the rare decays
  $B^+ \to \pi^+ \nu \bar\nu, B^+\to K^+ \nu\bar\nu$, and $B^+\to
  K^{*+}\nu\bar\nu$ in the Standard Model}},  {\em Phys. Lett. B} {\bf 680}
  (2009) 471--475, [\href{http://arxiv.org/abs/0908.1174}{{\tt
  arXiv:0908.1174}}].

\bibitem{Buras:2013ooa}
A.~J. Buras and J.~Girrbach, {\it {Towards the Identification of New Physics
  through Quark Flavour Violating Processes}},  {\em Rept.~Prog.~Phys.} {\bf
  77} (2014) 086201, [\href{http://arxiv.org/abs/1306.3775}{{\tt
  arXiv:1306.3775}}].

\bibitem{Buras:2015yca}
A.~J. Buras, D.~Buttazzo, and R.~Knegjens, {\it {$K\to\pi\nu\bar\nu$ and
  $\epsilon'/\epsilon$ in Simplified New Physics Models}},  {\em JHEP} {\bf 11}
  (2015) 166, [\href{http://arxiv.org/abs/1507.08672}{{\tt arXiv:1507.08672}}].

\bibitem{Buras:2022nfn}
A.~J. Buras, {\it {On the superiority of the $|V_{cb}|-\gamma $ plots over the
  unitarity triangle plots in the 2020s}},  {\em Eur. Phys. J. C} {\bf 82}
  (2022), no.~7 612, [\href{http://arxiv.org/abs/2204.10337}{{\tt
  arXiv:2204.10337}}].

\bibitem{King:2019lal}
D.~King, A.~Lenz, and T.~Rauh, {\it {$B_s$ mixing observables and
  $|V_{td}/V_{ts}|$ from sum rules}},  {\em JHEP} {\bf 05} (2019) 034,
  [\href{http://arxiv.org/abs/1904.00940}{{\tt arXiv:1904.00940}}].

\bibitem{UTfit:2022hsi}
{\bf UTfit} Collaboration, M.~Bona et~al., {\it {New UTfit Analysis of the
  Unitarity Triangle in the Cabibbo-Kobayashi-Maskawa scheme}},  {\em Rend.
  Lincei Sci. Fis. Nat.} {\bf 34} (2023) 37--57,
  [\href{http://arxiv.org/abs/2212.03894}{{\tt arXiv:2212.03894}}].

\bibitem{Brod:2013sga}
J.~Brod and J.~Zupan, {\it {The ultimate theoretical error on $\gamma$ from $B
  \to DK$ decays}},  {\em JHEP} {\bf 1401} (2014) 051,
  [\href{http://arxiv.org/abs/1308.5663}{{\tt arXiv:1308.5663}}].

\bibitem{Backus:2022xhi}
J.~V. Backus, M.~Freytsis, Y.~Grossman, S.~Schacht, and J.~Zupan, {\it {Toward
  extracting $\gamma$ from $B \to DK$ without binning}},
  \href{http://arxiv.org/abs/2211.05133}{{\tt arXiv:2211.05133}}.

\bibitem{Buras:2022nrb}
A.~J. Buras and E.~Venturini, {\it {Standard Model Predictions for Rare $K$ and
  $B$ Decays without $|V_{cb}|$ and $|V_{ub}|$ Uncertainties}},  3, 2022.
\newblock \href{http://arxiv.org/abs/2203.10099}{{\tt arXiv:2203.10099}}.

\bibitem{Straub:2015ica}
A.~Bharucha, D.~M. Straub, and R.~Zwicky, {\it {$B\to V\ell^+\ell^-$ in the
  Standard Model from light-cone sum rules}},  {\em JHEP} {\bf 08} (2016) 098,
  [\href{http://arxiv.org/abs/1503.05534}{{\tt arXiv:1503.05534}}].

\bibitem{Gubernari:2018wyi}
N.~Gubernari, A.~Kokulu, and D.~van Dyk, {\it {$B\to P$ and $B\to V$ Form
  Factors from $B$-Meson Light-Cone Sum Rules beyond Leading Twist}},  {\em
  JHEP} {\bf 01} (2019) 150, [\href{http://arxiv.org/abs/1811.00983}{{\tt
  arXiv:1811.00983}}].

\bibitem{Altmannshofer:2008dz}
W.~Altmannshofer, P.~Ball, A.~Bharucha, A.~J. Buras, D.~M. Straub, et~al., {\it
  {Symmetries and Asymmetries of $B \to K^{*} \mu^{+} \mu^{-}$ Decays in the
  Standard Model and Beyond}},  {\em JHEP} {\bf 0901} (2009) 019,
  [\href{http://arxiv.org/abs/0811.1214}{{\tt arXiv:0811.1214}}].

\bibitem{Descotes-Genon:2013vna}
S.~Descotes-Genon, T.~Hurth, J.~Matias, and J.~Virto, {\it {Optimizing the
  basis of ${B} \to {K}^{*}\ell^+ \ell^-$ observables in the full kinematic
  range}},  {\em JHEP} {\bf 1305} (2013) 137,
  [\href{http://arxiv.org/abs/1303.5794}{{\tt arXiv:1303.5794}}].

\bibitem{Bause:2021cna}
R.~Bause, H.~Gisbert, M.~Golz, and G.~Hiller, {\it {Interplay of dineutrino
  modes with semileptonic rare B-decays}},  {\em JHEP} {\bf 12} (2021) 061,
  [\href{http://arxiv.org/abs/2109.01675}{{\tt arXiv:2109.01675}}].

\bibitem{Li:2022xxs}
L.~Li, M.~Ruan, Y.~Wang, and Y.~Wang, {\it {Analysis of $B_s \rightarrow \phi
  \nu \bar\nu$ at CEPC}},  {\em Phys. Rev. D} {\bf 105} (2022), no.~11 114036.

\bibitem{Altmannshofer:2009ma}
W.~Altmannshofer, A.~J. Buras, D.~M. Straub, and M.~Wick, {\it {New strategies
  for New Physics search in $B \to K^{*} \nu \bar{\nu}$, $B \to K \nu
  \bar{\nu}$ and $B \to X_{s} \nu \bar{\nu}$ decays}},  {\em JHEP} {\bf 04}
  (2009) 022, [\href{http://arxiv.org/abs/0902.0160}{{\tt arXiv:0902.0160}}].

\bibitem{Buras:2014fpa}
A.~J. Buras, J.~Girrbach-Noe, C.~Niehoff, and D.~M. Straub, {\it {$B\to
  K^{(*)}\nu\bar\nu$ decays in the Standard Model and beyond}},  {\em JHEP}
  {\bf 1502} (2015) 184, [\href{http://arxiv.org/abs/1409.4557}{{\tt
  arXiv:1409.4557}}].

\bibitem{Kou:2018nap}
{\bf Belle-II} Collaboration, W.~Altmannshofer et~al., {\it {The Belle II
  Physics Book}},  {\em PTEP} {\bf 2019} (2019), no.~12 123C01,
  [\href{http://arxiv.org/abs/1808.10567}{{\tt arXiv:1808.10567}}]. [Erratum:
  PTEP 2020, 029201 (2020)].

\bibitem{Felkl:2021uxi}
T.~Felkl, S.~L. Li, and M.~A. Schmidt, {\it {A tale of invisibility:
  constraints on new physics in b \textrightarrow{}
  s\ensuremath{\nu}\ensuremath{\nu}}},  {\em JHEP} {\bf 12} (2021) 118,
  [\href{http://arxiv.org/abs/2111.04327}{{\tt arXiv:2111.04327}}].

\bibitem{Descotes-Genon:2020buf}
S.~Descotes-Genon, S.~Fajfer, J.~F. Kamenik, and M.~Novoa-Brunet, {\it
  {Implications of $b\to s\mu\mu$ anomalies for future measurements of $B \to
  K^{(*)} \nu \bar \nu$ and $K\to \pi \nu \bar \nu$}},  {\em Phys. Lett. B}
  {\bf 809} (2020) 135769, [\href{http://arxiv.org/abs/2005.03734}{{\tt
  arXiv:2005.03734}}].

\bibitem{He:2021yoz}
X.~G. He and G.~Valencia, {\it {$R^\nu(K^{(*)}$ and non-standard neutrino
  interactions}},  {\em Phys. Lett. B} {\bf 821} (2021) 136607,
  [\href{http://arxiv.org/abs/2108.05033}{{\tt arXiv:2108.05033}}].

\bibitem{Descotes-Genon:2021doz}
S.~Descotes-Genon, S.~Fajfer, J.~F. Kamenik, and M.~Novoa-Brunet, {\it
  {Implications of $b\to s\ell^+\ell^-$ constraints on $b\to s\nu\bar\nu$ and
  $s\to d\nu\bar\nu$}},  in {\em {55th Rencontres de Moriond on Electroweak
  Interactions and Unified Theories}}, 5, 2021.
\newblock \href{http://arxiv.org/abs/2105.09693}{{\tt arXiv:2105.09693}}.

\bibitem{Descotes-Genon:2022gcp}
S.~Descotes-Genon, S.~Fajfer, J.~F. Kamenik, and M.~Novoa-Brunet, {\it {Probing
  CP violation in exclusive $b \to s \nu \bar \nu$ transitions}},
  \href{http://arxiv.org/abs/2208.10880}{{\tt arXiv:2208.10880}}.

\bibitem{LHCb:2014cxe}
{\bf LHCb} Collaboration, R.~Aaij et~al., {\it {Differential branching
  fractions and isospin asymmetries of $B \to K^{(*)} \mu^+ \mu^-$ decays}},
  {\em JHEP} {\bf 06} (2014) 133, [\href{http://arxiv.org/abs/1403.8044}{{\tt
  arXiv:1403.8044}}].

\bibitem{LHCb:2016ykl}
{\bf LHCb} Collaboration, R.~Aaij et~al., {\it {Measurements of the S-wave
  fraction in $B^{0}\rightarrow K^{+}\pi^{-}\mu^{+}\mu^{-}$ decays and the
  $B^{0}\rightarrow K^{\ast}(892)^{0}\mu^{+}\mu^{-}$ differential branching
  fraction}},  {\em JHEP} {\bf 11} (2016) 047,
  [\href{http://arxiv.org/abs/1606.04731}{{\tt arXiv:1606.04731}}]. [Erratum:
  JHEP 04, 142 (2017)].

\bibitem{LHCb:2021zwz}
{\bf LHCb} Collaboration, R.~Aaij et~al., {\it {Branching Fraction Measurements
  of the Rare $B^0_s\rightarrow\phi\mu^+\mu^-$ and $B^0_s\rightarrow
  f_2^\prime(1525)\mu^+\mu^-$- Decays}},  {\em Phys. Rev. Lett.} {\bf 127}
  (2021), no.~15 151801, [\href{http://arxiv.org/abs/2105.14007}{{\tt
  arXiv:2105.14007}}].

\bibitem{LHCb:2015tgy}
{\bf LHCb} Collaboration, R.~Aaij et~al., {\it {Differential branching fraction
  and angular analysis of $\Lambda^{0}_{b} \rightarrow \Lambda \mu^+\mu^-$
  decays}},  {\em JHEP} {\bf 06} (2015) 115,
  [\href{http://arxiv.org/abs/1503.07138}{{\tt arXiv:1503.07138}}]. [Erratum:
  JHEP 09, 145 (2018)].

\bibitem{DELPHI:1996ohp}
{\bf DELPHI} Collaboration, W.~Adam et~al., {\it {Study of rare b decays with
  the DELPHI detector at LEP}},  {\em Z. Phys. C} {\bf 72} (1996) 207--220.

\bibitem{ALEPH:2000vvi}
{\bf ALEPH} Collaboration, R.~Barate et~al., {\it {Measurements of BR
  ($b\to\tau\bar\tau X)$ and BR ($b\to\tau\bar\tau D^{*\pm} X$ and upper limits
  on BR$ (B\to \tau\bar\tau$ and ) and BR $b\to s\nu\bar\nu)$}},  {\em Eur.
  Phys. J. C} {\bf 19} (2001) 213--227,
  [\href{http://arxiv.org/abs/hep-ex/0010022}{{\tt hep-ex/0010022}}].

\bibitem{Brod:2014bfa}
J.~Brod, A.~Lenz, G.~Tetlalmatzi-Xolocotzi, and M.~Wiebusch, {\it {New physics
  effects in tree-level decays and the precision in the determination of the
  quark mixing angle $\gamma$}},  {\em Phys.~Rev.} {\bf D92} (2015) 033002,
  [\href{http://arxiv.org/abs/1412.1446}{{\tt arXiv:1412.1446}}].

\bibitem{Lenz:2019lvd}
A.~Lenz and G.~Tetlalmatzi-Xolocotzi, {\it {Model-independent bounds on new
  physics effects in non-leptonic tree-level decays of B-mesons}},  {\em JHEP}
  {\bf 07} (2020) 177, [\href{http://arxiv.org/abs/1912.07621}{{\tt
  arXiv:1912.07621}}].

\bibitem{Iguro:2020ndk}
S.~Iguro and T.~Kitahara, {\it {Implications for new physics from a novel
  puzzle in $\bar{B}_{(s)}^0 \to D^{(\ast)+}_{(s)} \lbrace \pi^-, K^- \rbrace$
  decays}},  {\em Phys. Rev. D} {\bf 102} (2020), no.~7 071701,
  [\href{http://arxiv.org/abs/2008.01086}{{\tt arXiv:2008.01086}}].

\bibitem{Cai:2021mlt}
F.-M. Cai, W.-J. Deng, X.-Q. Li, and Y.-D. Yang, {\it {Probing new physics in
  class-I B-meson decays into heavy-light final states}},  {\em JHEP} {\bf 10}
  (2021) 235, [\href{http://arxiv.org/abs/2103.04138}{{\tt arXiv:2103.04138}}].

\bibitem{Bordone:2021cca}
M.~Bordone, A.~Greljo, and D.~Marzocca, {\it {Exploiting dijet resonance
  searches for flavor physics}},  {\em JHEP} {\bf 08} (2021) 036,
  [\href{http://arxiv.org/abs/2103.10332}{{\tt arXiv:2103.10332}}].

\bibitem{Lenz:2022kal}
A.~Lenz and S.~Monteil, {\it {High precision in CKM unitarity tests in $b$ and
  $c$ decays}},  7, 2022.
\newblock \href{http://arxiv.org/abs/2207.11055}{{\tt arXiv:2207.11055}}.

\bibitem{Faroughy:2022dyq}
D.~A. Faroughy, J.~F. Kamenik, M.~Szewc, and J.~Zupan, {\it {Accessing CKM
  suppressed top decays at the LHC}},
  \href{http://arxiv.org/abs/2209.01222}{{\tt arXiv:2209.01222}}.

\bibitem{DiLuzio:2017vat}
L.~Di~Luzio, A.~Greljo, and M.~Nardecchia, {\it {Gauge leptoquark as the origin
  of B-physics anomalies}},  {\em Phys. Rev. D} {\bf 96} (2017), no.~11 115011,
  [\href{http://arxiv.org/abs/1708.08450}{{\tt arXiv:1708.08450}}].

\bibitem{DiLuzio:2018zxy}
L.~Di~Luzio, J.~Fuentes-Martin, A.~Greljo, M.~Nardecchia, and S.~Renner, {\it
  {Maximal Flavour Violation: a Cabibbo mechanism for leptoquarks}},  {\em
  JHEP} {\bf 11} (2018) 081, [\href{http://arxiv.org/abs/1808.00942}{{\tt
  arXiv:1808.00942}}].

\bibitem{Cornella:2019hct}
C.~Cornella, J.~Fuentes-Martin, and G.~Isidori, {\it {Revisiting the vector
  leptoquark explanation of the B-physics anomalies}},  {\em JHEP} {\bf 07}
  (2019) 168, [\href{http://arxiv.org/abs/1903.11517}{{\tt arXiv:1903.11517}}].

\bibitem{Fuentes-Martin:2019ign}
J.~Fuentes-Mart\'\i{}n, G.~Isidori, M.~K\"onig, and N.~Selimovi\'c, {\it
  {Vector Leptoquarks Beyond Tree Level}},  {\em Phys. Rev. D} {\bf 101}
  (2020), no.~3 035024, [\href{http://arxiv.org/abs/1910.13474}{{\tt
  arXiv:1910.13474}}].

\bibitem{Fuentes-Martin:2020luw}
J.~Fuentes-Mart\'\i{}n, G.~Isidori, M.~K\"onig, and N.~Selimovi\'c, {\it
  {Vector leptoquarks beyond tree level. II. $\mathcal{O}(\alpha_s)$
  corrections and radial modes}},  {\em Phys. Rev. D} {\bf 102} (2020), no.~3
  035021, [\href{http://arxiv.org/abs/2006.16250}{{\tt arXiv:2006.16250}}].

\bibitem{Fuentes-Martin:2020hvc}
J.~Fuentes-Mart\'\i{}n, G.~Isidori, M.~K\"onig, and N.~Selimovi\'c, {\it
  {Vector Leptoquarks Beyond Tree Level III: Vector-like Fermions and
  Flavor-Changing Transitions}},  {\em Phys. Rev. D} {\bf 102} (2020) 115015,
  [\href{http://arxiv.org/abs/2009.11296}{{\tt arXiv:2009.11296}}].

\bibitem{Crosas:2022quq}
O.~L. Crosas, G.~Isidori, J.~M. Lizana, N.~Selimovic, and B.~A. Stefanek, {\it
  {Flavor Non-universal Vector Leptoquark Imprints in $K\to \pi \nu\bar \nu$
  and $\Delta F = 2$ Transitions}},  {\em Phys. Lett. B} {\bf 835} (2022)
  137525, [\href{http://arxiv.org/abs/2207.00018}{{\tt arXiv:2207.00018}}].

\bibitem{Aoki:2021kgd}
Y.~Aoki et~al., {\it {FLAG Review 2021}},
  \href{http://arxiv.org/abs/2111.09849}{{\tt arXiv:2111.09849}}.

\bibitem{Brod:2021hsj}
J.~Brod, M.~Gorbahn, and E.~Stamou, {\it {Updated Standard Model Prediction for
  $K \to \pi \nu \bar{\nu}$ and $\epsilon_K$}},  {\em PoS} {\bf BEAUTY2020}
  (2021) 056, [\href{http://arxiv.org/abs/2105.02868}{{\tt arXiv:2105.02868}}].

\bibitem{Buras:2010pza}
A.~J. Buras, D.~Guadagnoli, and G.~Isidori, {\it {On $\epsilon_K$ beyond lowest
  order in the Operator Product Expansion}},  {\em Phys.~Lett.} {\bf B688}
  (2010) 309--313, [\href{http://arxiv.org/abs/1002.3612}{{\tt
  arXiv:1002.3612}}].

\bibitem{Urban:1997gw}
J.~Urban, F.~Krauss, U.~Jentschura, and G.~Soff, {\it {Next-to-leading order
  QCD corrections for the $B^0 - \bar B^0$ mixing with an extended Higgs
  sector}},  {\em Nucl.~Phys.} {\bf B523} (1998) 40--58,
  [\href{http://arxiv.org/abs/hep-ph/9710245}{{\tt hep-ph/9710245}}].

\bibitem{Amhis:2016xyh}
{\bf Heavy Flavor Averaging Group (HFAG)} Collaboration, Y.~Amhis et~al., {\it
  {Averages of $b$-hadron, $c$-hadron, and $\tau$-lepton properties as of
  summer 2016}},  \href{http://arxiv.org/abs/1612.07233}{{\tt
  arXiv:1612.07233}}.

\bibitem{Aoki:2019cca}
{\bf Flavour Lattice Averaging Group} Collaboration, S.~Aoki et~al., {\it {FLAG
  Review 2019: Flavour Lattice Averaging Group (FLAG)}},  {\em Eur. Phys. J. C}
  {\bf 80} (2020), no.~2 113, [\href{http://arxiv.org/abs/1902.08191}{{\tt
  arXiv:1902.08191}}].

\end{thebibliography}\endgroup

\end{document}